\documentclass[]{mn2e}
\usepackage{psfig, epsf, epsfig}
\title[Abundance inhomogeneity]{Origin of Abundance Inhomogeneity
in Globular Clusters}
\author[K. Bekki, S. W. Campbell, J. C. Lattanzio, J. E. Norris]
       {K. Bekki${}^1$\thanks{E-mail: bekki@bat.phys.unsw.edu.au},
   S. W. Campbell${}^2$, J. C. Lattanzio${}^2$,  and J. E. Norris${}^3$ \\
        ${}^1$School of Physics, University of New South Wales,
              Sydney 2052, NSW, Australia \\
        ${}^2$ Centre for Stellar and Planetary Astrophysics, School of Mathematical Sciences,  Monash University,  Australia \\
        ${}^3$ Research School of Astronomy \& Astrophysics,
        The Australian National University,
        Mt Stromlo Observatory,
        Australia}

\begin{document}

\date{Accepted, Received 2005 May 13; in original form }

\pagerange{\pageref{firstpage}--\pageref{lastpage}} \pubyear{2005}

\maketitle

\label{firstpage}

\begin{abstract}

We numerically investigate abundance properties of 
the Galactic globular clusters
(GCs) by adopting  a new ``external pollution'' scenario.
In this framework,
GCs  are assumed to originate  in forming low-mass dwarfs embedded in
dark matter subhalos  at very high redshifts ($z$)
and thus be  chemically influenced by field AGB stars 
of the dwarfs during early GC formation processes.
GCs within a dwarf galaxy  therefore
can be formed from the mixture of (i) gas ejected from 
the field
AGB stars formed earlier in the dwarf 
and (ii) the interstellar gas infalling to the central
region of the dwarf.
In this external  pollution scenario, 
the ratio of the total mass of infalling gas
to that of AGB ejecta during GC formation in a dwarf 
($s$) and the time scale of 
gas infall (${\sigma}_{\rm I}$) are the most important key parameters  
that can determine abundance properties of GCs.
We mainly investigate the abundance inhomogeneity  among
light elements (e.g., C, N, O, Na, and Al) of stars in GCs 
by using the latest stellar yield models of metal-poor AGB stars 
{\it with
and without third dredge-up.}
Our principal results for the models with no third dredge-up,
which are  more consistent with observations, 
are as follows. 

(1) Both [N/Fe] and [C/Fe] can be diverse among stars
within a GC owing to chemical pollution from field AGB stars. 
[N/Fe] distributions in some GCs can clearly show 
bimodality whereas [C/Fe] is  monomodal in most models.
[N/Fe] distributions depend on $s$ such that
models with smaller $s$ (i.e., larger mass fraction
of AGB ejecta used for GC formation)  show 
the [N/Fe] bimodality more clearly.

(2) N-rich, C-poor stars in GCs also have higher He abundances
owing to pollution from massive AGB stars with He-rich ejecta.
The number fraction of He-rich stars (Y $>0.30$) is higher
for the models with smaller $s$ 
and shorter ${\sigma}_{\rm I}$
for $3\le s \le24$ and $10^5 \le {\sigma}_{\rm I} \le 10^7$ yr.
He abundances of stars  correlate with [N/Fe] and [Al/Fe]
and anticorrelate with [C/Fe],  [O/Fe], and  [Na/Fe] 
within GCs in our models.

(3) Although our model can much better explain the observed
C-N and Mg-Al anticorrelations than previous theoretical models,
it is in strong disagreement with the observed O-Na anticorrelation.

(4) This model naturally provides 
an explanation for the large fraction of 
CN-strong stars without recourse to an implausible IMF.

Based on these results for  the above external pollution scenario, 
we discuss the long-standing problem
of the CN-bimodality prevalent in the Galactic GCs,
the possible helium abundance inhomogeneity in these systems,
and their horizontal branch morphologies.

\end{abstract}


\begin{table*}
\centering
\begin{minipage}{185mm}
%\begin{minipage}
\caption{Stellar yields of AGB stars}
\begin{tabular}{cccccccccccc}
model                                 &
H  & 
He                         & 
C              &  
N                                    & 
O                      &  
Na &
$^{24}$Mg  & 
$^{25}$Mg  & 
$^{26}$Mg  & 
Al  &
Fe  \\
YA1 & 
.6639E+0 &
.3345E+0 &
.5316E-4 &
.1143E-2 &
.1592E-4 &
.4365E-6 &
.2806E-6 &
.2786E-4 &
.3429E-5 &
.1855E-4 &
.5481E-4  \\
YA2 & 
.6414E+0 &
.3491E+0 &
.1206E-2 &
.7358E-2 &
.2475E-3 &
.1438E-4 &
.9788E-6 &
.8031E-4 &
.1121E-3 &
.1067E-4 &
.5375E-4 \\
\end{tabular}
\end{minipage}
{\bf Notes:} The mass fraction of AGB ejecta is given for each element.
These yields are for AGB stars with masses of $6.5 {\rm M}_{\odot}$
and [Fe/H] = $-1.4$.
%The H mass fraction is 0.6639 for YA1 and  0.6414 for YA2 and
%not listed for conveneice (due to the width limitaion of the table).
Stellar yields for the AGB stars with differences masses and metallicities
are from  Campbell et al. (2006).
\end{table*}

\begin{table*}
\centering
\begin{minipage}{185mm}
%\begin{minipage}
\caption{Initial abundances of infalling gas}
\begin{tabular}{cccccccc}
model                                 &
Y  & 
[C/Fe]                         & 
[N/Fe]              &  
[O/Fe]              &  
[Na/Fe]                                    & 
[Mg/Fe]                      &  
[Al/Fe]   \\ 
YI1 & 0.23 &  0.0 &  -0.25 & 0.3 & -0.3  & 0.5 & 0.1  \\
YI2 & 0.23 &  1.25 & -1.97 & -1.63 & -0.3  & 0.38 & -1.74 \\
YI3 & 0.23 &  0.0 &  -0.25 & 0.3 & -0.3  & 0.5 & 0.3  \\
YI4 & 0.23 &  -0.3 &  -0.25 & 0.3 & -0.3  & 0.5 & 0.1  \\
\end{tabular}
\end{minipage}
\end{table*}

\begin{table*}
\centering
\begin{minipage}{185mm}
\caption{Model parameters}
\begin{tabular}{ccccccc}
(1)&(2)&(3)&(4)&(5)&(6)&(7)\\
Model no.                                  &
Gas yield  &  
AGB yield                        & 
$s$ ($M_{\rm IN}/M_{\rm AGB}$)               &  
${\sigma}_{\rm I}$ (in units of $10^7$ yr)                         & 
$M_{\rm g}(0)$   (in units of $M_{\rm f}$)                &
Comments  \\
M1 & YI1 &  YA1 & 3.0 & 0.1  & 0.01 & The fiducial model \\
M2 & YI2 &  YA1 & 3.0 & 0.1  & 0.01 &  \\
M3 & YI3 &  YA1 & 3.0 & 0.1  & 0.01 &  \\
M4 & YI1 &  YA2 & 3.0 & 0.1  & 0.01 &  with third dredge-up\\
M5 & YI1 &  YA1 & 6.0 & 0.1  & 0.01 &  \\
M6 & YI1 &  YA1 & 12.0 & 0.1  & 0.01 &  \\
M7 & YI1 &  YA1 & 24.0 & 0.1  & 0.01 &  \\
M8 & YI1 &  YA1 & 3.0 & 0.01  & 0.01 &  \\
M9 & YI1 &  YA1 & 3.0 & 0.05  & 0.01 &  \\
M10 & YI1 &  YA1 & 3.0 & 1.0  & 0.01 &  \\
M11 & YI1 &  YA1 & 3.0 & 0.1  & 0.001 & A model for NGC 6752 \\
M12 & YI1 &  YA1 & 3.0 & 0.1  & 0.005 &  \\
M13 & YI1 &  YA1 & 3.0 & 0.1  & 0.029 &  \\
M14 & YI1 &  YA1 & 0.17 & 0.01  & 0.001 & A model for $\omega$ Cen \\
M15 & YI4 &  YA1 & 3.0 & 0.1  & 0.01 & smaller initial [C/Fe] \\
\end{tabular}
\end{minipage}
{\bf Notes:} Cols. (2--3) yield types for infalling gas and AGB ejecta,
(shown in tables 1 and 2) 
respectively. Cols. (4--6) model parameters.
\end{table*}

\section{Introduction}

Since observational evidence of star-to-star abundance inhomogeneity
among the 
 light elements of stars in the Galactic globular clusters (GCs)
was discovered  
(e.g., Cohen 1978; Peterson 1980; Norris et al. 1981; Leep et al. 1986), 
the origin of the inhomogeneity has been extensively discussed
both theoretically and observationally (e.g.,  Sneden et al. 1992;
Norris \& Da Costa 1995; Smith et al. 2005; 
see Gratton et al. 2004 for a recent review).
So far the following two scenarios (or working hypotheses)
have been proposed 
for the  origin of the abundance inhomogeneity:
(1) the primordial hypothesis and (2) the mixing hypothesis 
(e.g., Sweigart \& Mengel 1979;
Cottrell \& Da Costa 1981; Freeman \& Norris 1981;
Smith 1987; Suntzeff 1993; Kraft 1994; Denissenkov \& Weiss 1996).
The first hypothesis is that the observed inhomogeneity 
is due to the second generation of stars that 
formed from  gas ejected from
the first generation of evolved stars (e.g., AGB stars)
(``primordial pollution'' or 
``self-pollution''  scenarios, e.g.,  Cottrell \& Da Costa 1981).

Internal processes of stars,
such as dredge-up of the  CN-processing of  envelope  material in 
inner hydrogen-burning regions, 
are key for understanding the observed  chemical inhomogeneity of GCs
in the second mixing hypothesis
(e.g., 
``deep mixing''; Smith 1987; Kraft 1994; Thoul et al. 2002).
Although deep mixing is highly unlikely
to occur in  less evolved stars
on the main sequence and subgiant-branch,
star-to-star  abundance inhomogeneity 
was found in these stars
of some Galactic GCs
(e.g., Cannon et al. 1998 for 47 Tuc).
The  self-pollution scenario is accordingly suggested
to be more promising
than the mixing (evolutionary) scenario
(Da Costa et al. 2004; Gratton 2004 for a recent review).

Stellar ejecta from AGB stars and massive stars have been considered 
to play a key role in early chemical evolution of GCs for
the self-pollution scenario (e.g., Fenner et al. 2004; 
Charbonnel \& Prantzos 2006; Prantzos \& Charbonnel 2006;
Karakas et al. 2006; Bekki \& Chiba 2007). 
Previous studies suggested that GCs are unlikely to retain
AGB ejecta owing to ram pressure stripping by warm/cold  gas
of the Galactic halo and disk (e.g., Frank \& Gisler 1976;
Smith 1996; Gnedin et al. 2002).
These imply that  GCs
are  even more unlikely to  retain AGB ejecta
in protogalactic environments, where denser halo/disk gas
can strip the AGB ejecta efficiently from GCs during hydrodynamical
interaction between the gas and the AGB ejecta.
These previous studies thus appear to suggest
that self-pollution is {\it not} likely 
within  GCs evolving in isolation,
though  numerical attempts 
have not yet been made to investigate 
the details of self-pollution processes
within GCs.

Furthermore, the self-pollution scenario is suggested to have
the following two  problems in explaining physical
properties of GCs. 
The first is that
the observed large number fraction of CN-strong populations
($\sim 0.5$ for NGC6752),
which correspond to the second generation of stars formed
from AGB ejecta, can not be explained by the scenario
without assuming very unusual and unrealistic IMFs
(e.g., Smith \& Norris 1982; D'Antona and Caloi 2004): 
the total mass of AGB ejecta
from the first generation of stars in a GC is
only $1-10$\% of the total mass of the GC with
a canonical IMF  (Bekki \& Norris 2006).
The second is that the observed O$-$Na and Mg$-$Al anticorrelations 
and C-depleted stars in CN-strong populations can not be reproduced
quantitatively
by chemical evolution models based on the self-pollution scenario
(Fenner et al. 2004). This problem, however, could  be due to 
the adopted AGB models, which have some uncertainties in predicting
chemical yields (e.g., 
Denissenkov \& Herwig 2003;
Karakas \& Lattanzio 2003; Ventura \& D'Antona 2005a,b; 
Campbell et al. 2006). 
%The third is that AGB ejecta can be hardly retained in
%GCs {\it evolving in isolation} if they are influenced by
%ram pressure stripping of intergalactic warm halo gas and
%interstellar disk gas of the Galaxy (e.g., Frank \& Gisler 1976;
%Smith 1996).

The most serious problem among these  is the first
one -- the observed large fraction of stars 
that (in this scenario) formed 
from AGB ejecta.
The very  large mass fraction of ``polluted'' second
generation of stars in the self-pollution scenario 
requires that 
AGB ejecta used for the formation of this
generation of stars originate from stellar systems that
had  total masses much larger than those of the present GCs
and have already disappeared for some reason. 
There could  be two possible scenarios that can satisfy the above
requirement.
In the first place, we propose:
(i) the GCs were initially 
more massive than the present-day GCs  by a factor of $10-100$,
(ii) AGB ejecta of the first generation of stars were
consumed by the formation of  the second generation
with very high star formation efficiency,
and (iii) most ($90-99$\%) of the first generation of stars had been 
{\it preferentially}  lost
by some physical mechanisms such as tidal stripping due
to the Galactic tidal field.
This scenario can be viable if enough  of the
AGB ejecta can be converted into the second  generation in
less than $10^8$ yr after the completion of 
the first generation Type II supernovae
(SNeII) -- otherwise, age differences
between the first and the second generation would  cause a
large  age spread inconsistent with the observed tightness of
the color-magnitude diagrams  of GCs.

In the second  scenario,  proposed by
Bekki (2006):
(i) proto-globular clouds are located 
in the central regions of  low-mass proto-galaxies
embedded in dark matter subhalos,
(ii)  AGB ejecta of the host galaxy's field stars,
which  formed earlier
and surround the clouds, 
goes into star formation within  the clouds
on  a time scale of $\sim 10^7$ yr (i.e., until the intracluster gas 
is expelled by SNeII),
and (iii) the low-mass galaxies were destroyed 
by the Galactic tidal fields to
become the Galactic halo components  whereas the GCs
were tidally stripped to form the Galactic halo GCs. 
In this scenario  there can be very small age differences
($\sim  10^7$yr) between stars with different abundances
in a single GC. 
This pollution process would be better called ``external pollution''
rather than ``self-pollution'',  because most of AGB ejecta 
originates from {\it outside} the  GCs.
Although Bekki \& Norris (2006) have recently discussed
the origin of the possible helium overabundance
of stellar populations of $\omega$ Cen in the context
of the external pollution scenario,
the abundance inhomogeneities observed in  ``normal'' GCs have not been
extensively studied using  previous chemical evolution
models based on the {\it external pollution} scenario.

The purpose of this paper  is thus to discuss the origin
of abundance inhomogeneity observed in the lighter elements
(e.g., C, N, O, Na, and Al) of GCs by
using chemical evolution models based on the external
pollution scenario. 
In particular, we 
discuss
(1) the [C/Fe]-[N/Fe] anticorrelation,
(2) CN-bimodality, (3) the O$-$Na and Mg$-$Al anticorrelations,
%(3) abundace ratios of ${}^{25} {\rm Mg}/{}^{24} {\rm Mg}$
%and ${}^{26} {\rm Mg}/{}^{24} {\rm Mg}$ and their correlations
%with O, Na, Mg, and Al abundances, 
(4) the helium abundance, 
and (5) observable predictions
of the models.
We adopt the AGB models recently developed by Campbell et al. (2006)
in which the third dredge up in low-mass AGB stars can,
in effect,  be switched off and on 
due to the use or non-use of the Ledoux criterion for 
convective boundaries.
We mainly show the results of the models {\it without} the third dredge-up,
because the models  can better explain the observed
abundance inhomogeneity if they are included in the present
chemical evolution models.
We realize that these models are ad-hoc, but they enable us to
investigate the proposed formation mechanism. 
% The plan of the paper is as follows: in the next section,
%we describe the external pollution scenario and the adopted
%chemical evolution models. 
%In \S 3, we present the numerical results
%on the abundance properties of GCs -- in particular on
%CN-bimodality and  the O$-$Na and Mg$-$Al anticorrelations. 
%In \S 4, we discuss advantages and disadvantages of
%the external pollution scenario in explaining
%chemical properties of GCs. 
%We summarise our  conclusions in \S 5.
%Since many authors have already discussed advantages and disadvantages
%of the primordial pollution (or enrichment) processes {\it within
%isolated proto-GC clouds and  GCs}
%(e.g., Cayrel 1986; 
%Parmentier \& Gilmore 2001;
%Thoul et al. 2002;
%Denissenkov \& Herwig 2003;
%Recchi \& Danziger 2005), 
%here we  do not intend to discuss these points.

\section{The model}

\subsection{The external pollution scenario}

We consider GC formation in the central regions of low-mass, proto-galaxies
embedded in dark matter halos at very  high  $z$.
The GC host galaxies can provide a deep gravitational potential
that allows AGB ejecta to be retained and used for star formation
in proto-GC clouds.
 In this scenario,
both (i) AGB ejecta from field stellar populations (i.e., the main components 
of the galaxies)
and (ii) proto-galactic  interstellar gas (which falls into the central regions)
can be mixed with each other and subsequently used for GC formation. 
Gas ejected from SNeII with velocities of $\sim 1000$ km s$^{-1}$
is assumed to be expelled from dwarfs during galaxy formation
and thus irresponsible for the chemical evolution of forming GCs.
After GC formation,  the host galaxies merge with the proto-Galaxy and
are completely destroyed by the strong Galactic tidal field.
The field stellar components of the galaxies
are dispersed into the Galactic halo
region to become old, metal-poor halo stars whereas GCs,
which are not destroyed by the Galaxy owing to their  initial compactness,
become the Galactic halo GCs.
%The schematic diagram and several advantages of this scenario
%are given in the  Appendix A.

Star formation and chemical evolution are assumed to proceed
until type II supernovae (SNe II) formed within proto-GCs
prevent further star formation   in this scenario.
Therefore, star formation is assumed to continue for less than $\sim 10^7$ yr 
in the models in which stars with different masses form in
a coeval way.
Age differences of stars are typically  less than
5 Myr for most models in the present study.
Evolution of chemical abundances depends mainly on (1) how
much  AGB ejecta is mixed into fresh ISM and consequently
converted into stars  
and (2) the time scale of gas infall into central regions of galaxies.
Advantages and disadvantages  of the external pollution
scenario are discussed later in \S 4.1.

\begin{figure}
\psfig{file=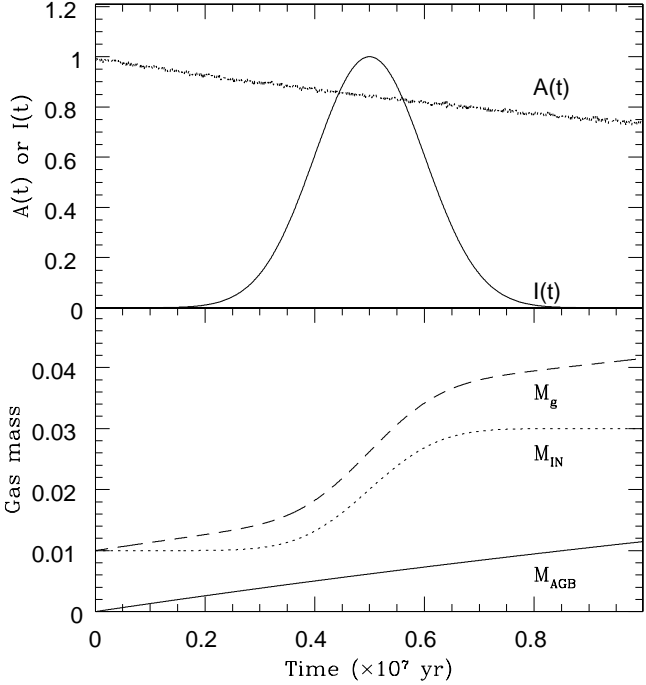,width=8.5cm}
\caption{ 
The upper panel shows the
time  evolution of the gas infall rate ($I(t)$, solid)
and the gas ejection rate of AGB stars ($A(t)$, dotted)
in the fiducial model M1.
The lower panel shows the time evolution of
the total mass of AGB ejecta 
accumulated in the  proto-GC cloud ($M_{\rm AGB}$, solid),
the  gas accreted in  the cloud via gas infall ($M_{\rm IN}$, dotted), 
and the  gas in  the cloud  ($M_{\rm g}$, dashed)
for M1 without star formation.
The total mass of the proto-GC cloud 
increases both by accretion of AGB ejecta and by gas infall.
}
\label{Figure. 1} 
\end{figure}

\subsection{Star formation and gas mass evolution}

The total gas mass ($M_{\rm g}(t)$)
of a forming GC changes through the gas ejection
of AGB stars, gas infall, and star formation:
\begin{equation}
\frac{dM_{\rm g}}{dt}=A(t)+I(t)-S(t),
\end{equation}
where  $A(t)$, $I(t)$, and $S(t)$ are the gas ejection rate
of the field AGB stars, gas infall rate, and star formation rate,
respectively. In order to calculate $A(t)$ at each time step 
for the  field stars (which have  a  total mass of $M_{\rm f}$),
we assume an
IMF in number defined
as $\psi (m_{\rm I}) = A_{0}{m_{\rm I}}^{-q}$,
where $m_{\rm I}$ is the initial mass of
each individual star and the slope $q=2.35$ corresponds to the Salpeter IMF.
The normalization factor $A_{0}$ is a function of $M_{\rm f}$,
$m_{\rm l}$ (lower mass cut-off), and $m_{\rm u}$ (upper mass cut-off):
\begin{equation}
A_{0}=\frac{M_{\rm f} \times (2-q)}{{m_{\rm u}}^{2-q}-{m_{\rm l}}^{2-q}}.
\end{equation}
where $m_{\rm l}$ and $m_{\rm u}$ are  set to be   $0.1 {\rm M}_{\odot}$
and  $120 {\rm M}_{\odot}$, respectively.
We adopt $q=2.35$ for all models in the present study.
$A(t)$ between $t$ and $t+dt$ is accordingly described as:
\begin{equation}
A(t)=\frac{1}{dt} \times \int_{m_{1}}^{m_{2}} f_{\rm ej} 
\psi (m) dm,
\end{equation}
where $m_{1}$ and  $m_{2}$ are masses of
stars that turn off the main sequence (thus becomes a AGB star)
at the time $t+dt$ and $t$, respectively (thus $m_{1} < m_{2}$).
$f_{\rm ej}$ describes the total gas mass ejected from  
an  AGB star with initial mass  $m_{\rm I}$ 
and final mass  ($m_{\rm F}$).
We derive an analytic form of $f_{\rm ej}$
($=m_{\rm I}-m_{\rm F}$) from  the observational data
by Weidemann (2000) by using the least-square fitting method, and  
find:
\begin{equation}
f_{\rm ej} =0.916 M_{\rm I}-0.444.
\end{equation}

In order to calculate the main-sequence
turn-off mass ($m_{\rm TO} =  m_1$ and $m_2$) 
at each time step, we use the following formula
(Renzini \& Buzzoni 1986; Norman \& Scoville 1988):
\begin{equation} 
\log m_{\rm TO}(t_{\rm s})
= 0.0558 (\log t_{\rm s})^2 - 1.338 \log t_{\rm s} + 7.764,
\end{equation} 
where $m_{\rm TO}$  is in solar units and time $t_{\rm s}$ in years.
We assume that the time $t=0$ corresponds to the epoch ($T_0$) when
the most massive AGB star (with $m_{\rm I}=8 {\rm M}_{\odot}$)
starts to eject gas in the present models. 
Therefore $t_{\rm s}=t+T_0$ in the above equation (5). 
 
The gas infall rate $I(t)$ is described as 
\begin{equation}
I(t)=I_{0} \exp ( \frac{-{(t-T_{\rm I})}^{2}}{2 {{\sigma}_{\rm I}}^2}), 
\end{equation}
where $I_{0}$ is the normalization factor for the total mass of
infalling gas, $T_{\rm I}$ represents the epoch of the maximum
infall rate, and ${\sigma}_{\rm I}$ describes the time-scale of
gas infall. The adopted Gaussian distribution means that
the gas infall rate increases, whether slowly or  rapidly,  -- depending
on the adopted  ${\sigma}_{\rm I}$, for $0 \le t \le T_{\rm I}$ --
and then decreases for  $T_{\rm I} < t$. 

The star formation  rate $S(t)$ is described as 
\begin{equation}
S(t)=S_{0} {M_{\rm g}}^{\alpha} 
\end{equation}
where $S_{0}$ (a normalization constant) is determined for 
a given  $\alpha$ so that most of the gas can be  consumed 
within an order of  $\sim 10^7$ yr. 
Although we investigate
a range of $\alpha$, we show the results for the models
with $\alpha=1$ which can better 
explain observations.

Observations (e.g., tightness of color-magnitude  diagrams of GCs)
strongly suggest that 
the timescale of GC  formation should be  quite short (an order of $10^8$ yr).
We also consider that star formation is truncated owing
to feedback effects of Type-II supernovae (SNe II) formed during   
GC formation:
the present models with this truncation can keep abundance homogeneity
in heavy elements (e.g., Fe) and thus can be consistent
with physical properties of normal GCs in the Galaxy. 
Therefore  we assume that gas can be converted
into stars in GCs only  for $0\le t \le T_{\rm end}$,
where $T_{\rm end}$ should be similar to or
shorter than $10^7$ yr corresponding
to the typical lifetimes of stars that  explode as SNe II. 
We present the results of the representative 
models with $T_{\rm end}=10^7$ yr
and $T_{\rm I}/T_{\rm end}=0.5$  in the present study.
The gas mass ($M_{\rm g}$) is given in units of
$M_{\rm f}$ (the total mass of field stars).
In the adopted IMF,
$M_{\rm AGB}$ is 0.01 ($ \times M_{\rm f}$) 
at $t=T_{\rm end}$ for all models.

\subsection{Abundance evolution}
The time evolution of the abundance of the $i$-th element
is described as follows:
\begin{equation}
\frac{d(Z_{i}M_{\rm g})}{dt}=Z_{{\rm A},i}(t)A(t)
+Z_{{\rm I},i}(t)I(t)
-Z_{i}S(t),
\end{equation}
where $Z_{{\rm A},i}(t)$ and 
 $Z_{{\rm I},i}(t)$  are the $i$-th abundance 
of AGB ejecta and infalling gas, respectively.
$Z_{{\rm A},i}(t)$  is time-dependent,  because 
of the varying timescales of evolution  for the different
masses of AGB stars (see  equation (5)). 
We assume that $Z_{{\rm I},i}(t)$  is  constant  for each $i$-th element
and thus refer to it  simply as $Z_{{\rm I},i}$
in the present study.

In order to calculate $Z_{{\rm A},i}(t)$ , we use the latest chemical yield
tables produced by Campbell et al. (2006) in which
H, He, C, N, O, Na, ${}^{24}$Mg, ${}^{25}$Mg, ${}^{26}$Mg,
Al, are listed for AGB stars with the masses ranging
from $2.5 {\rm M}_{\odot}$ to   $6.5 {\rm M}_{\odot}$ for a given [Fe/H]
($-1.4$).
These models follow normal 
AGB evolution with hot bottom burning, 
but we produce two sets of yields - one with the normal third dredge-up
and one
without (see below for details).
The chemical yield model without (with) the third dredge-up
are referred to as YA1 (YA2) in the present study.
Table 1 shows the mass fraction of the above eleven species 
in AGB ejecta from Campbell et al. (2006).
We consider that  $Z_{{\rm I},i}$ should be similar to typical
values of old halo stars in the Galaxy 
(e.g., Goswami \& Prantzos 2000; Venn et al. 2004) 
and the adopted values of $Z_{{\rm I},i}$
are summarized in Table 2 for four different models.
The details of the models without third dredge-up
are given in Campbell (2006).
Throughout this study, we assume that 
[Fe/H] is the same between infalling gas and AGB ejecta.
This assumption is reasonable, because we consider that gas 
from SNeII can be expelled from dwarf efficiently during galaxy formation.
We do not include the effects of massive rotating stars
on the chemical evolution of proto-GC clouds in the present study,
because
we intend to understand more clearly 
the roles of AGB stars in the chemical evolution:
we need  to derive separately the roles of AGB stars and those
of massive rotating stars in the chemical evolution.

\subsection{With and without third dredge-up}

We mainly show the results of  the models that
use the AGB yields with  no third dredge-up,
though we investigate the models with and without third dredge-up.
It should be stressed here that
the models with no third dredge-up are
purely hypothetical ones that have not
yet been proved.
Previous studies based on  self-pollution models
with third dredge-up in AGB stars failed to explain many of
the observations of abundance inhomogeneity in GCs
(e.g., Fenner et al. 2004).
If our models can  successfully reproduce some
observations that were not explained by previous ones,
we can obtain some insights into the ``culprits'' 
in the self-pollution or external pollution scenarios. 
The abundance inhomogeneity of  GCs in 
the present models with no third dredge-up
originates  from pure hydrogen
burning, 
which is available in HBB AGB stars.

The present models with third dredge-up
can not explain at all most of the observed correlations
between different abundances (e.g., the [C/Fe]-[N/Fe] anticorrelation).
For example, these models can not produce  stars with [C/Fe] $<0$ and 
[N/Fe] $>0$, because [N/Fe] increases with the increase of [C/Fe]
during the chemical evolution owing to the C-rich ejecta 
of AGB stars. Furthermore, the observed CN-bimodality of
stars can not be reproduced in these models, owing to the 
lack of C-depleted, N-rich stars in these models.
These failures would suggest that {\it
AGB stars are not the polluters that are responsible for 
the observed abundance inhomogeneity in GCs}.
We however  consider that it is still worthwhile to investigate
abundance properties of GCs by using different AGB models (i.e.,
those with no third dredge-up) in the present study
for the above-mentioned reason.
It should also be  stressed here that 
the present models with no third dredge-up can better 
explain observations,  because we adopt the star formation
histories of proto-GC clouds in the external pollution scenario:
successful reproduction of some observations in the present
study is due to the {\it combination  of the adopted models
both with no third dredge-up and star formation histories 
of the external pollution scenario}.

Further, at present there are no yields available for the
so-called "Super-AGB Stars", which are more massive
AGB stars that experience thermal pulses after core
carbon burning. The few calculations in existence
(e.g.,  Ritossa et al  1996; Gil-Pons et al 2005; Siess 2006)
show that these stars experience very hot HBB and
very little dredge-up. Qualitatively this is what is
required to solve the GC problem, so our models
without dredge-up may simulate to some extent the
yields from these stars, while we await full evolution
and nucleosynthesis calculations for the appropriate
masses and compositions.

We close this section with some comments
about the reliability of AGB star yields.
We believe that the balance of evidence
indicates that AGB stars with dredge-up are
not the site of the abundance inhomogeneities
seen in globular cluster stars. However, there
are still so many uncertainties that it may
still be possible for these stars to produce the
abundance profiles required. For example, the
depth of dredge-up is still very uncertain, the
temperature profile in the HBB envelope depends
on the convection theory (e.g., Ventura \& D'Antona 2005a, b, c) 
as well as associated parameters,
mass-loss and of course various reaction rates.
Although previous studies (Ventura et al. 2002; Ventura \& D'Antona 2005c)
provided useful tables for stellar yields of AGB stars,
the lack of tables for some abundances
(e.g., the lack of a table for Al abundance
in  Ventura \& D'Antona 2005c)
currently does not allow us to investigate the O-Na and Mg-Al anticorrelations
in a fully self-consistent manner. 
It is thus our future study
to compare the present results with those based on chemical
evolution models using tables from works other than
Campbell (2006).  
More comments on AGB models with and without third dredge-up
are given in the Appendix B.

\subsection{Main points of analysis}

One of the most important parameters that govern the chemical
evolution of GCs in the external pollution scenario is the
ratio ($s$) 
of the total mass of the infalling gas ($M_{\rm IN}$) to that
of AGB ejecta ($M_{\rm AGB}$).
We investigate the models with  $s$ ($M_{\rm IN}/M_{\rm AGB}$)
ranging from 0.1 to 100 and thereby isolated the  best values
of $s$ that can explain observations. 
The  time scale of gas infall (${\sigma}_{\rm I}$)
is also an important parameter that
can control final abundance distributions of GCs in the 
present study. We investigate  models
with ${\sigma}_{\rm I}$  ranging from $10^5$ yr to
$10^7$ yr and show important results of the models.

Although these $s$ and ${\sigma}_{\rm I}$ are the most important
parameters, the initial gas mass $M_{\rm g}(0)$ can also
change abundance distributions in a less dramatic manner
for a given $s$ and ${\sigma}_{\rm I}$. We investigate the models with
$0.001 \le M_{\rm g}(0) \le 0.029$ and show the results of some
representative models.
Table 3 summarizes the parameter values used for  the
15 models investigated in the present study.
We first present the results of the fiducial model M1,
which shows interesting and important results, 
in \S 3.1.
We then describe the dependences of the results on
the two key parameters ($s$ and ${\sigma}_{\rm I}$)
in \S 3.2. 

By changing $s$, ${\sigma}_{\rm I}$, and $M_{\rm g}(0)$, we can construct
a model that can explain reasonably well the abundance distribution
for a specific GC. For example,  the model  M11  shows a bimodal
[N/Fe] distribution with the normalized number fraction of N-rich
and C-depleted stars being equal to that of the N-normal and C-normal
ones and thus is consistent with the observed CN distribution in 
NGC 6752 (e.g., Smith \& Norris 1993).
Since we are focusing  on the general behaviors 
resulting from  parameter dependences 
of the models, we do not discuss extensively the
abundance distributions of individual GCs. 
We however just briefly discuss the origin of He abundance distributions
of different stellar populations in $\omega$ Cen
in the Appendix A,
because this issue is quite topical and is currently being
investigated  extensively by 
different theoretical models 
(e.g., Bekki \& Norris 2006;  Maeder \& Meynet 2006). 

In order to compare the present models 
with previous ones (e.g., Fenner et al. 2004, Ventura \& D'Antona 2005c),
we compare the present results with the same observational ones 
(Yong et al. 2003 for NGC 6752)
that have been already  compared also with these previous studies
in \S 3.
We also compare the present results 
with other  more recent   observations
on abundances of stars (in particular, less evolved stars) in GCs
(e.g. Ramirez \& Cohen 2002;  Briley et al. 2004a,b; 
Carretta et al. 2005; Cohen et al. 2005)
and assess  the viability of the external pollution scenario
later in \S 4.1

\begin{figure}
\psfig{file=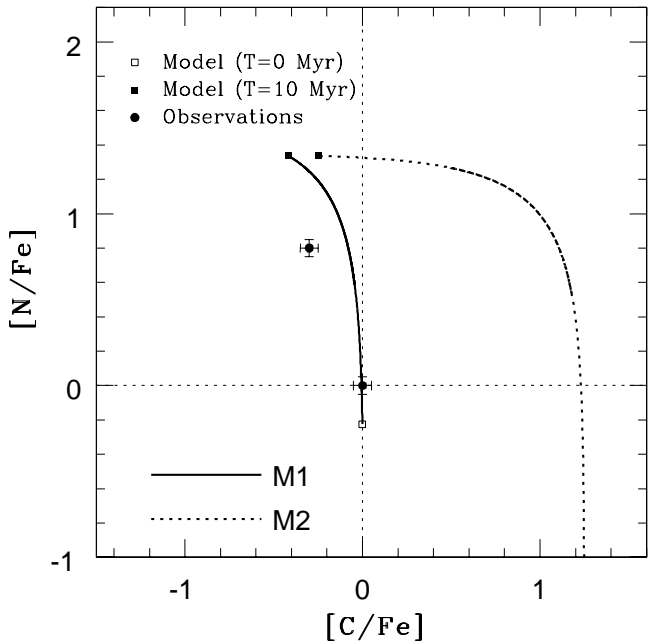,width=8.5cm}
\caption{ 
Time evolution of the fiducial model M1 (solid)
and M2 (dotted) on the [C/Fe]$-$[N/Fe]
plane. The initial and final values are plotted
by open and filled squares, respectively.
For comparison, observational results by 
Smith \& Norris (1993) for  AGB stars
in NGC 6752 are shown by filled circles.
}
\label{Figure. 2} 
\end{figure}

\begin{figure}
\psfig{file=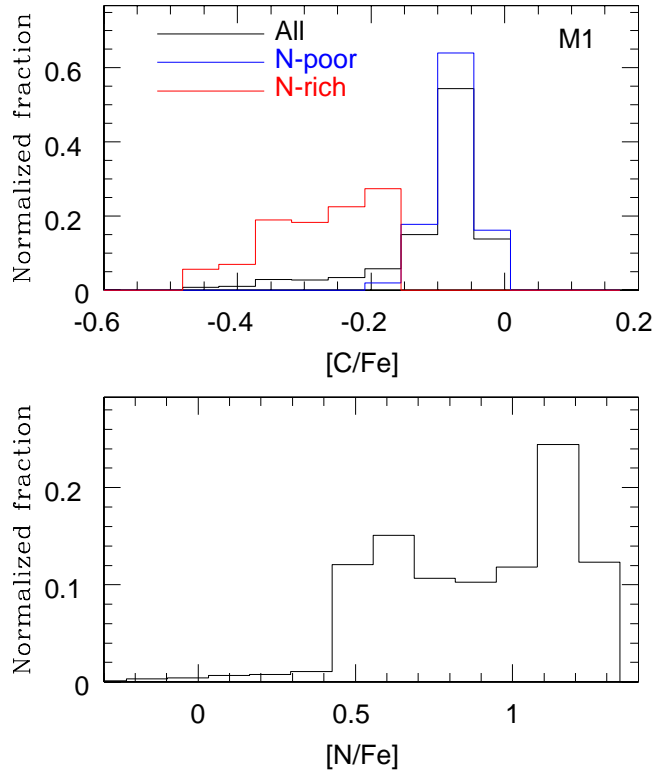,width=8.5cm}
\caption{ 
Normalized number distributions of [C/Fe] (upper)
and [N/Fe] (lower) in the fiducial model M1.
The [C/Fe] distributions are shown  separately for
N-poor stars with [N/Fe] $<1$ (blue)
and for N-rich ones with [N/Fe] $\ge 1$ (red)
so that a clear difference in the [C/Fe] distributions
between the two populations can be seen.
}
\label{Figure. 3} 
\end{figure}

\section{Results}

\subsection{The fiducial model}

\subsubsection{C and N abundances}

Fig. 2 shows the time evolution of the fiducial model M1 
in  comparison with the  model M2 on the [C/Fe]-[N/Fe] plane. 
There are differences in initial abundances of infalling gas between
models M1 and M2 (e.g., [C/Fe] and [N/Fe]).
Since these
abundances are estimated for new stars formed at each time
step in a GC, each location of the line represents
abundances of stars with given ages 
rather than the mean abundances of the GC. 
As the proto-GC cloud is polluted by ejecta
from more massive field AGB stars ($m_{\rm I}> 5 {\rm M}_{\odot}$), [N/Fe] 
and [C/Fe] in the cloud rapidly increases and decreases, respectively. 
Gas infall increases the total gas mass of the cloud (and thus 
the density) and consequently  increases gradually 
star formation in the cloud. As a result of this,
new stars that formed from the gas,
which is a mix of  infalling gas and AGB ejecta,
can have higher [N/Fe] and lower [C/Fe]. As a larger amount of
AGB ejecta is accumulated in the cloud at later times,
new stars formed later can have higher [N/Fe] and lower [C/Fe]
than those formed earlier.
Thus stars with higher  [N/Fe] can show lower [C/Fe] in
the fiducial model M1 and this anticorrelation between
[N/Fe] and  [C/Fe] can be seen in most models.

Although the fiducial model can qualitatively explain
the observed anticorrelation in  Fig. 2 (e.g., Norris et al. 1981;
Smith \& Norris 1993),
it does not pass the observed location of the N-rich population 
in NGC 6752. This quantitative inconsistency between the model
and the observation can be seen in the comparative model M2
with initially small [N/Fe] in the  infalling gas that can significantly
dilute  [N/Fe] during star formation within GCs. 
In the present model,  [N/Fe] always becomes significantly high ($\sim 1$)
{\it when [C/Fe] becomes low irrespective of the infall rate
of N-poor gas.}
Since only two data points (i.e., two stars) shown in Fig. 2
are obviously not enough to make any robust conclusions
on the quantitative consistency of the present model with observations,
we discuss the  consistency
of the model  by using more data sets later in \S 4.1.

\begin{figure}
\psfig{file=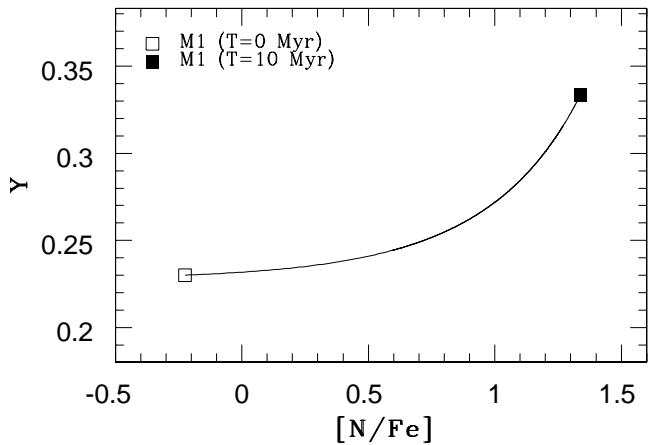,width=8.5cm}
\caption{ 
The same as Figure 2 but for Y (helium abundance by  mass)
and [N/Fe] for M1.
}
\label{Figure. 4} 
\end{figure}

\begin{figure}
\psfig{file=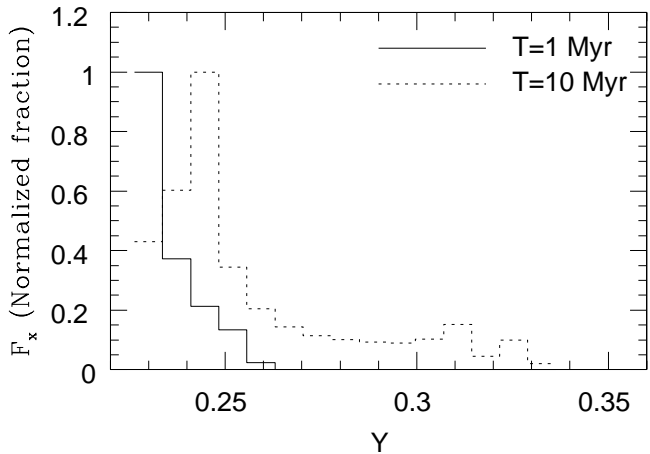,width=8.5cm}
\caption{ 
The distribution of Y (helium abundance) 
normalized to the maximum values (referred to as $F_{\rm x}$
in what follows) at the time T 
= 1 Myr (solid) and 10 Myr (dotted) for  M1.
}
\label{Figure. 5} 
\end{figure}

\begin{figure}
\psfig{file=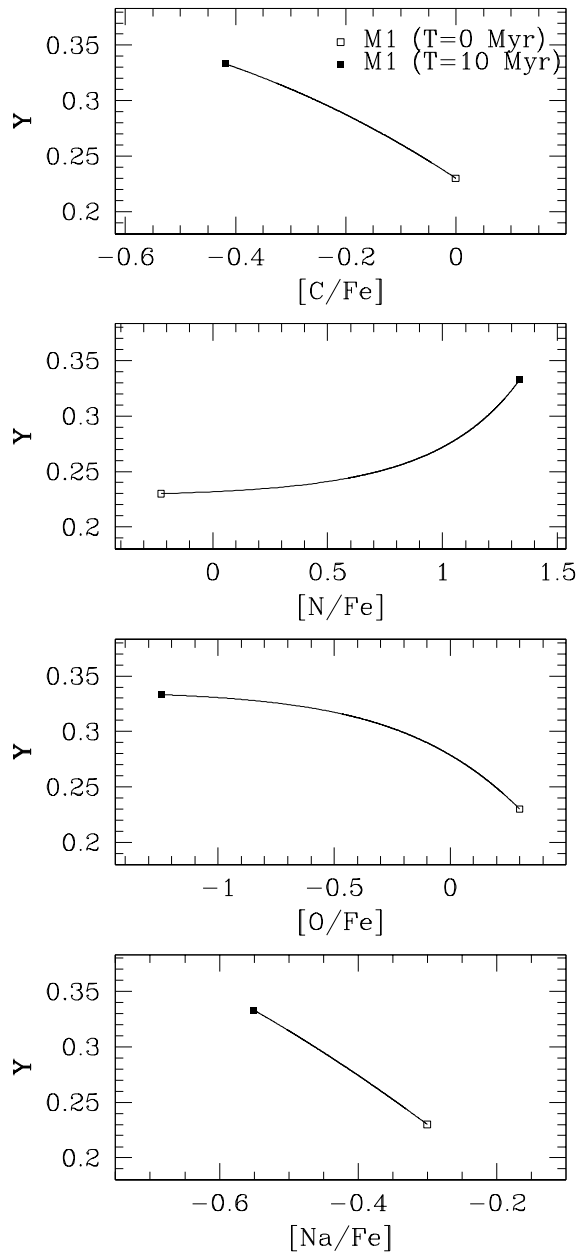,width=7.5cm}
\caption{ 
The same as Figure 4 but for [C/Fe]-Y (top),
[N/Fe]-Y (second from top),
[O/Fe]-Y (second from bottom),
and [Na/Fe]-Y (bottom) for M1.
}
\label{Figure. 6} 
\end{figure}

As a result of gradual increase (decrease)
of [N/Fe] ([C/Fe]) due to AGB pollution,
new stars formed at different times can have
different [N/Fe] and [C/Fe] and thus show 
abundance inhomogeneity in  [N/Fe] and [C/Fe].
Fig. 3, showing the normalized number distributions of stars  
with different [N/Fe] and [C/Fe] in M1,
demonstrates that the [N/Fe] distribution has a clearer bimodality
than the [C/Fe] distribution. 
Most N-rich stars with [N/Fe] $>1$
show a lower  [C/Fe] ($<-0.3$), as expected from chemical pollution
from massive AGB stars. 
Owing to the adopted density-dependent star formation rate,
the gas infall  causes two peaks in the star formation rate in
M1. New stars formed during the first peak of star formation
can have lower [N/Fe] whereas those formed during the second peak
can have higher [N/Fe] due to a higher degree of AGB pollution. 
The reason for the apparent absence of the bimodality in [C/Fe] is that
[C/Fe] can not dramatically change within a short time scale
($\sim 10^7$ yr)
so that new stars can show a much narrower [C/Fe] distribution
(i.e., the bimodality can not be evident for the adopted binning).
The final {\it mean}  [C/Fe] 
and [N/Fe] at $T=10^7$ yr are -0.09 and 0.85, respectively.

\begin{figure}
\psfig{file=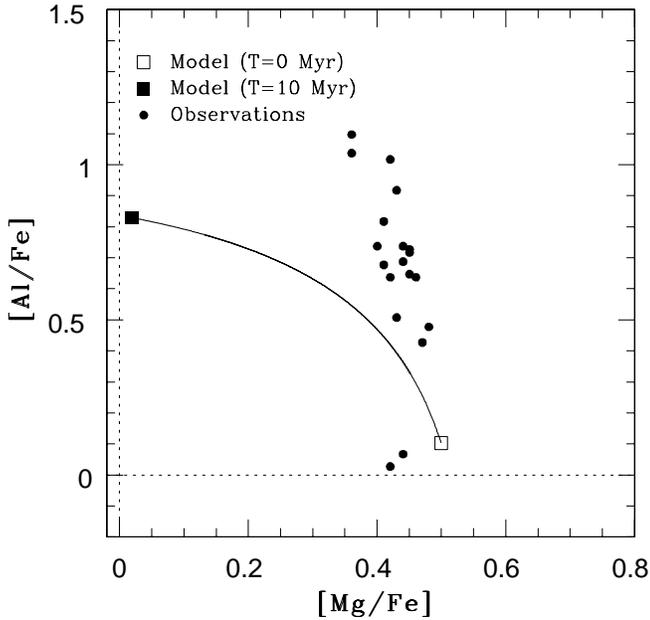,width=8.5cm}
\caption{ 
The same as Figure 2 but for the [Mg/Fe]-[Al/Fe] relation
for M1. For comparison, observations from Yong et al. (2003)
are shown by filled circles.
The original values of abundances listed  in Yong et al. (2003)
are used for the observational plots.
}
\label{Figure. 7} 
\end{figure}

\begin{figure}
\psfig{file=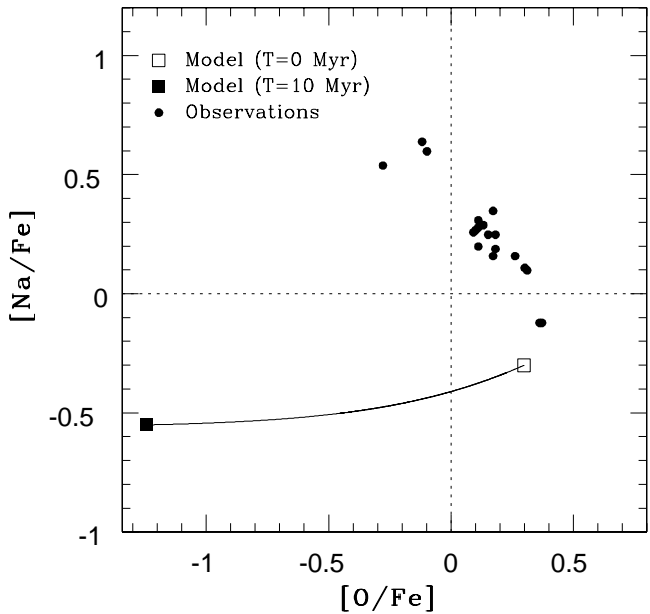,width=8.5cm}
\caption{ 
The same as Figure 2 but for the [O/Fe]-[Na/Fe] relation
for M1. For comparison, observations from Yong et al. (2003)
are shown by filled circles.
}
\label{Figure. 8} 
\end{figure}
\subsubsection{He abundance}

As the gas mass of the proto-GC cloud  increases 
owing to accumulation of AGB material in M1,
the He abundance (Y) of the gas  also increases
in a monotonic fashion:
the final {\it mean} Y 
at $T=10^7$ yr is 0.26 whereas
the final Y of stars {\it formed at  $T=10^7$ yr} is 0.33.
As a result of this,  new stars formed later will  have
higher Y than those formed earlier.
Since AGB ejecta increases [N/Fe] rapidly,
new stars that form later and thus have higher Y 
can  show higher [N/Fe]. 
Fig. 4 clearly shows a positive yet 
non-linear correlation between Y and [N/Fe] in new stars of
GCs for M1. This correlation can be seen in almost all
models in the present study and thus is simply
a result of H-burning.

Owing to the time-evolution of Y in the gas cloud,
new stars formed at different times show different
Y,  which results in a Y inhomogeneity in the GC.
Fig. 5 clearly shows that as the proto-GC cloud 
is polluted by AGB ejecta to a larger extent,
newly formed stars show a larger degree of  Y inhomogeneity.
Furthermore, the location of the peak in the Y distribution is also shifted
to the higher values as the AGB pollution proceeds:
the peak value changes from 0.23 at $T=1$ Myr to 0.25 at $T=10$ Myr.
These results imply that the time scale of GC formation is 
an important factor which can determine Y distributions of GCs.

Since abundances in the present model either increase or decrease
in a monotonic fashion as AGB pollution proceeds,
strong correlations or anticorrelations can be expected between
different abundances of  stars in a GC.
Fig. 6 shows that Y can anticorrelate with [C/Fe], [O/Fe],
and [Na/Fe],  though the dependences are not necessarily linear.
These correlations can be expected for any chemical evolution
models of GCs using stellar yields of AGB stars with no 3rd
dredge-up.

\begin{figure}
\psfig{file=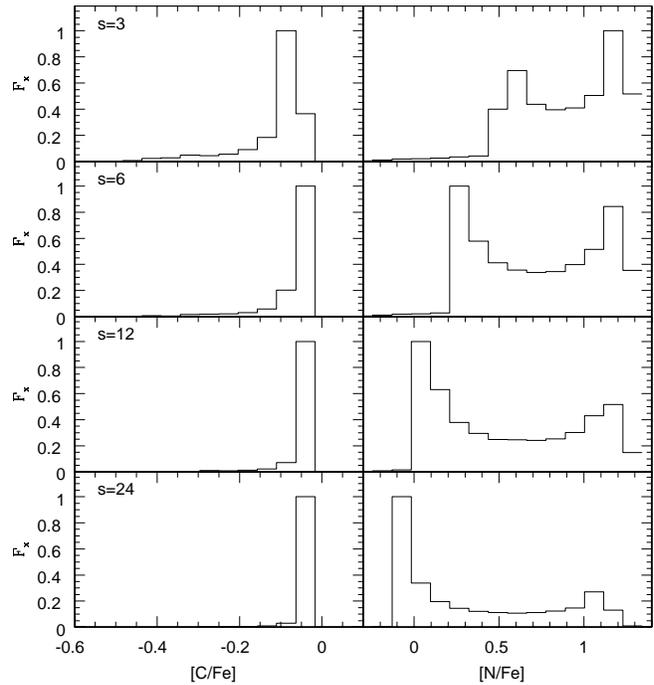,width=8.5cm}
\caption{ 
The dependences of [C/Fe] (left four) and 
[N/Fe] (right four) distributions ($F_{\rm x}$)
on $s$. 
}
\label{Figure. 9} 
\end{figure}

\begin{figure}
\psfig{file=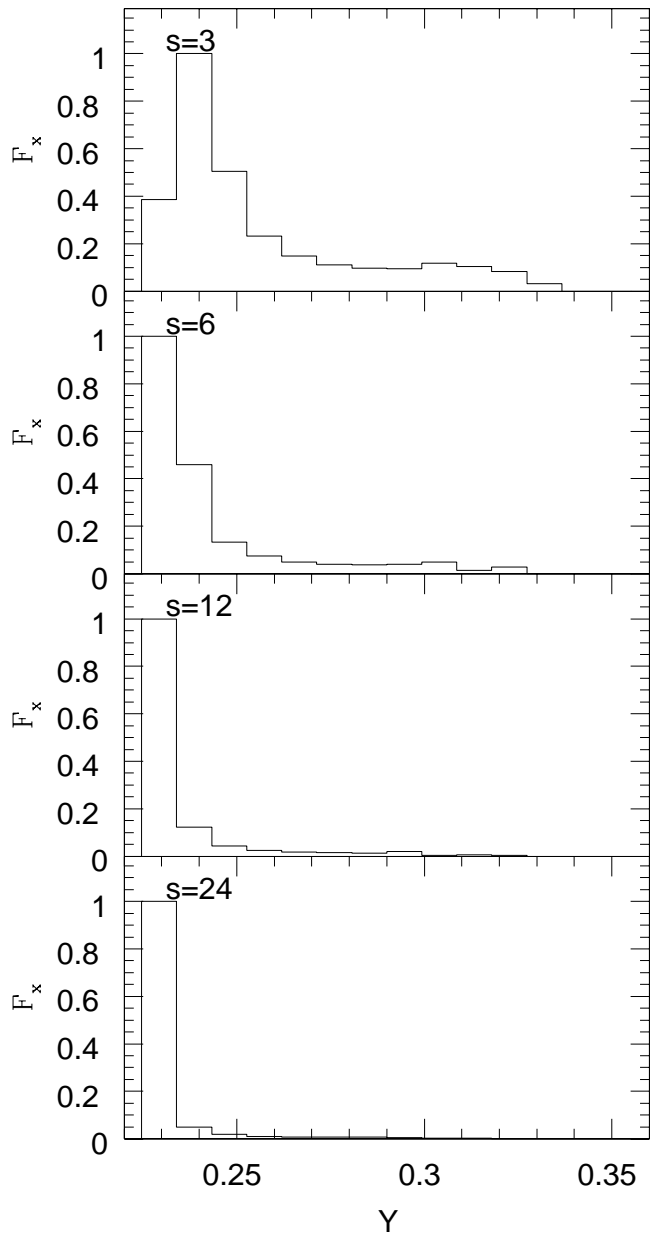,width=8.5cm}
\caption{ 
The dependences of Y 
distributions ($F_{\rm x}$)
on $s$. 
}
\label{Figure. 10} 
\end{figure}

\begin{figure}
\psfig{file=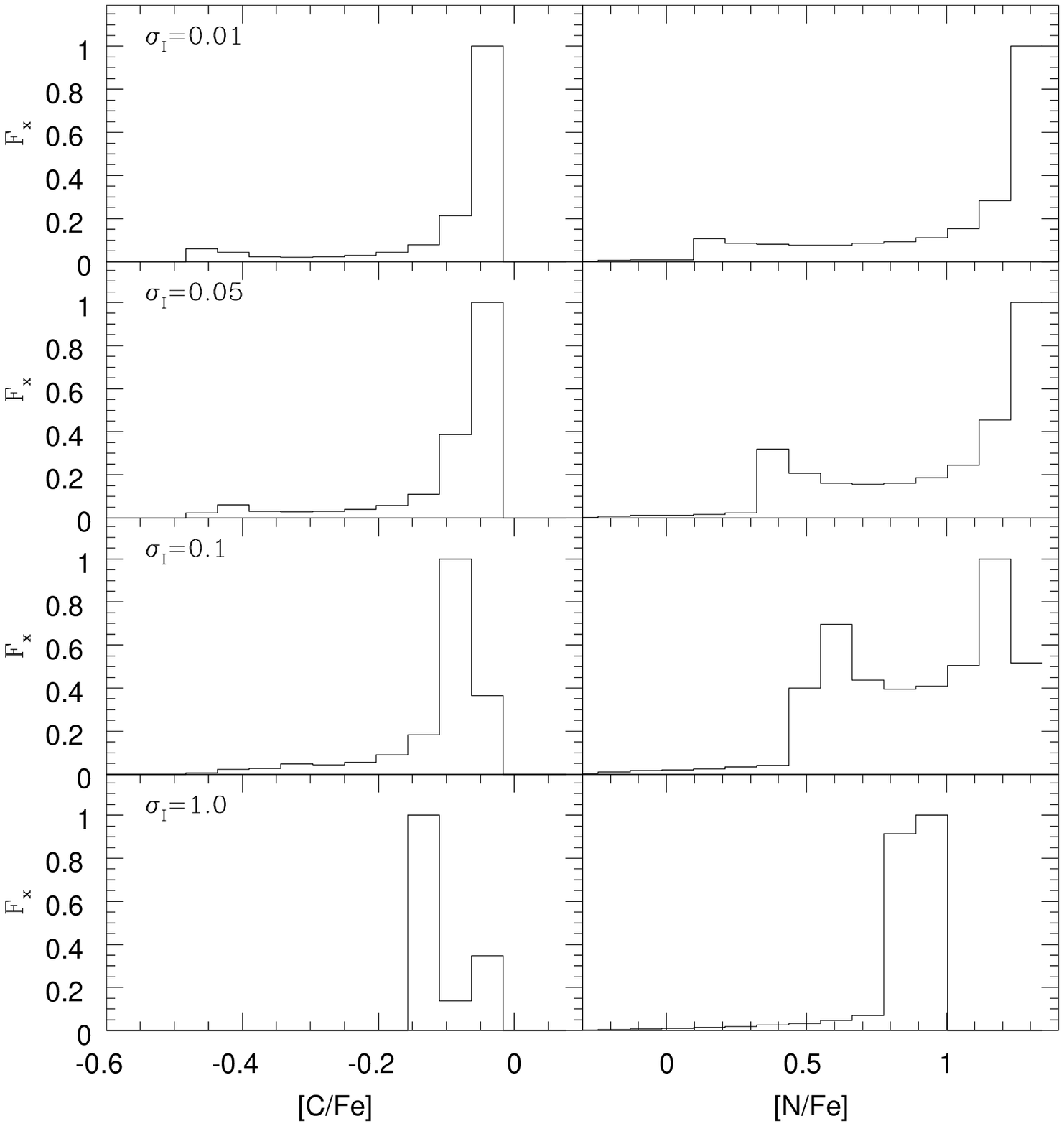,width=8.5cm}
\caption{ 
The same as Figure 9 but as a function of ${\sigma}_{\rm I}$.
}
\label{Figure. 11} 
\end{figure}

\begin{figure}
\psfig{file=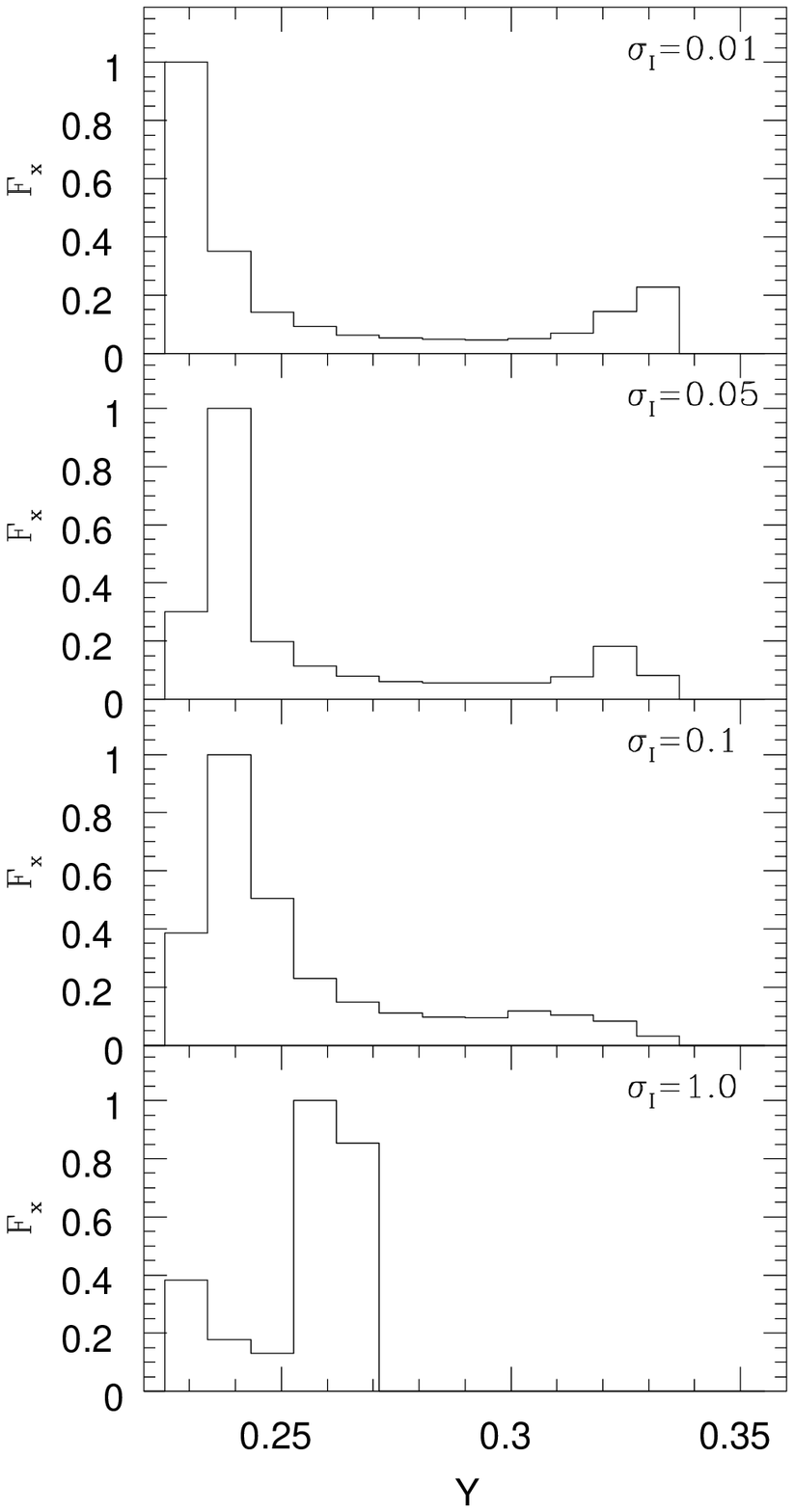,width=8.5cm}
\caption{ 
The same as Figure 10 but as a function of  ${\sigma}_{\rm I}$.
}
\label{Figure. 12} 
\end{figure}
\subsubsection{Mg-Al and Na-O anticorrelations}

Fig. 7 shows the time evolution of M1 in the [Mg/Fe]-[Al/Fe]
plane and the observations of bright red giant stars
in NGC 6752 (Yong et al. 2003). 
As a natural result of ongoing chemical pollution of the proto-GC
cloud by field AGB stars, 
the [Mg/Fe] and [Al/Fe] ratios  of new stars within the GC
decrease and increase, respectively, as AGB pollution
proceeds.
Although the observed  [Mg/Fe]-[Al/Fe] anticorrelation
can be well reproduced qualitatively, the observed slope of
the anticorrelation and   the ranges of [Mg/Fe] and [Al/Fe] 
are not consistent with the predicted values: a shallower slope
of the anticorrelation, 
a wider range of [Mg/Fe], a smaller range of [Al/Fe] are clearly seen
in the model. This inconsistency can be seen in the model
M3 with initially higher [Al/Fe].
Although all models in the present study show  
[Mg/Fe]-[Al/Fe] anticorrelations, none of the models shows
slopes as steep as the observed one for NGC 6752.

Fig. 8 shows the time evolution of M1 in  the [O/Fe]-[Na/Fe]
plane as well as  the observational plots of bright red giant stars
in NGC 6752.
The model predicts a positive correlation between [O/Fe]-[Na/Fe],
which is obviously inconsistent with the observed 
[O/Fe]-[Na/Fe] anticorrelation. Furthermore, the predicted
[Na/Fe] is too low to match the observed values,
and this inconsistency can be seen in all models of the 
present study.
This suggests that the derived inconsistency is due to the
adopted AGB yield models in which no third dredge-up is
assumed. 
Also the present study  with [O/Fe]-[Na/Fe] correlation
is in contrast with 
Fenner et al. (2004) in which third dredge-up
is assumed and the derived [O/Fe]-[Na/Fe]  anticorrelation
is not quantitatively consistent with the observed one
 by Yong et al. (2003).
The fact that both Fenner et al. (2004)  and the present study
fail to explain the observed  [O/Fe]-[Na/Fe] anticorrelation quantitatively
may well suggest that AGB pollution scenarios have  a serious
problem in terms of explaining the observed 
abundance inhomogeneity in GCs.
We will discuss this point later in \S 4.1.

\subsection{Parameter dependences}

\subsubsection{The dependence on $s$}

Fig. 9 shows the dependences of [C/Fe] and [N/Fe] number distributions
normalized to their maximum values (denoted as $F_{\rm x}$) on $s$, i.e.,  
the ratio of $M_{\rm IN}$ (the total mass of infalling gas)
to $M_{\rm AGB}$ (the total mass of AGB ejecta) and thus
controls the degree of AGB pollution.
Fig. 9 clearly shows that both the shapes and the peak values
of the distributions depend on $s$, which suggests that
$s$ is an important parameter for [C/Fe] and [N/Fe] distributions.
Bimodal distributions can be seen in [N/Fe] for models 
with $3 \le s \le 24$ 
and the fraction of N-rich stars with [N/Fe] $>1$ is 
higher in the models with smaller $s$ (i.e., larger mass
fraction of AGB ejecta).
As $s$ becomes larger,
both the difference between smaller and larger peaks in [N/Fe]
and $F_{\rm x}$ in the smaller peak
become  larger.
The derived diverse distributions of [N/Fe]
can be compared with the observed diversity 
in the CN distributions of GCs (Norris 1988).

The [C/Fe] distributions, on the other hand, do not depend so strongly
on $s$ and the four models show almost monomodal distributions.
This reflects the fact that chemical pollution by AGB stars
does not change [C/Fe] as dramatically as [N/Fe].
The spread in [C/Fe] is however dependent on $s$ in the sense
that the spread is larger for models with smaller $s$
(i.e., a larger degree of AGB pollution).
The predicted bimodal [N/Fe] distribution and monomodal
[C/Fe] one are discussed later in \S 4.1 by comparing
these predictions with the latest observations for less-evolved
stars (e.g. MS stars).

Fig. 10 shows that Y distributions normalized
to the maximum values ($F_{\rm x}$) depend on $s$
such that both the peak values and the spreads in the Y distributions
are larger for the models with smaller $s$ 
(a larger degree of AGB pollution).
These results accordingly suggest  that GCs with larger degrees of 
Y inhomogeneity will  show higher Y. 
The results shown in Figs. 9 and 10 therefore
suggest that GCs with higher fractions of 
N-rich stars with [N/Fe] $>1$ can show higher Y and
larger degrees of Y  inhomogeneity.
Since the fraction of N-rich stars in a GC can be observationally
quantified by  an  ``$r$'' parameter by counting  
the number fraction of  CN-strong stars
vs CN-weak stars  (Norris 1988), 
the above predicted correlation can be tested
by investigating correlations between the $r$ parameters
and observable properties  dependent on
Y  inhomogeneity, such as color spreads between 
MS stars in the color-magnitude  diagrams of GCs.

\subsubsection{The dependence on ${\sigma}_{\rm I}$}

Fig. 11 shows that [N/Fe] distributions depend strongly
on ${\sigma}_{\rm I}$, which means that [N/Fe] distributions can be controlled
by the time-scale of gas infall.
As seen in the $s$ dependences of [N/Fe],
the models with higher peak values of [N/Fe] show larger 
degrees of [N/Fe] spreads in the distributions. 
It is significant that the model with ${\sigma}_{\rm I}=1.0$
shows an almost monomodal [N/Fe] distribution with an
apparently small dispersion of [N/Fe] (i.e. no clear bimodality). 
[C/Fe] distributions do not depend so strongly on ${\sigma}_{\rm I}$ 
as [N/Fe] so that models show a relatively narrow range
of [C/Fe].
It can be concluded from Figs. 9 and 11 that
(i) [C/Fe] distributions are highly likely to be monomodal
irrespective of $s$ and ${\sigma}_{\rm I}$
and (ii) [N/Fe] distributions are likely to be bimodal
and it depends on  $s$ and ${\sigma}_{\rm I}$ how clearly
the distributions show the bimodality.

Fig. 12 shows that the models with smaller  ${\sigma}_{\rm I}$
(i.e., more rapid gas infall) show larger degrees of 
Y  inhomogeneity. However 
a positive correlation between the peak  Y 
and the degrees of Y inhomogeneity seen in 
models with different $s$ (shown in Fig. 10) 
can not be seen in these models with different ${\sigma}_{\rm I}$.
Given the fact that mean Y should not be 
different between models with the same $M_{\rm AGB}$ 
(or $s$) yet different ${\sigma}_{\rm I}$,
$r$ parameters can be correlated {\it only} with
the degrees of Y inhomogeneity (i.e., not with Y).
The results shown in Figs. 10 and 12 thus suggest that 
Y distributions are  diverse depending on the two 
key parameters $s$ and ${\sigma}_{\rm I}$.
Observational implications of these results 
are discussed later in \S 4.4.

Thus the present study
shows that Y, [C/Fe], and [N/Fe] distributions
in GCs are predicted to be  diverse if GCs are formed from 
the mixture of infalling gas and AGB ejecta of field
AGB stars in forming dwarfs at high $z$.
The results shown in Figs. 10-12 can provide
some physical basis not only for the  observed
diversity in CN distributions of GCs (e.g., Norris 1988)
but also for the possible Y spreads in NGC 2808 
(e.g., D'Antona et al. 2005) and $\omega$ Cen (e.g., Bedin et al. 2004).
It should be stressed here that the diverse distributions
come from
{\it the combination of the external pollution scenario and
the adopted AGB yield}: neither the external pollution scenario
assuming the third dredge-up in AGB stars nor 
classical self-pollution scenarios assuming no third dredge-up
are able to predict  diverse distributions similar
to the observed ones.

%\subsubsection{[N/Fe] and [C/Fe] distributions}

\begin{figure}
\psfig{file=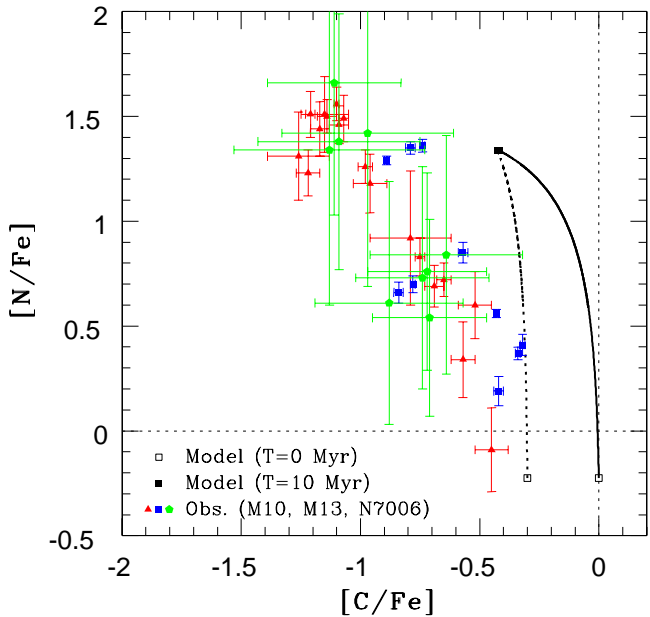,width=8.5cm}
\caption{ 
Comparison between models (M1 and M15) 
and observations (Smith et al. 2005) for M10  (red triangles),
M13 (blue squares), and NGC 7006 (green pentagons)   
in the [C/Fe]-[N/Fe] plane.
The time evolution of M1 and M15 are shown by
thick solid and dotted lines, respectively.
The observational data are for red giants in these
GCs (i.e., for more evolved stars). 
}
\label{Figure. 13} 
\end{figure}

\begin{figure}
\psfig{file=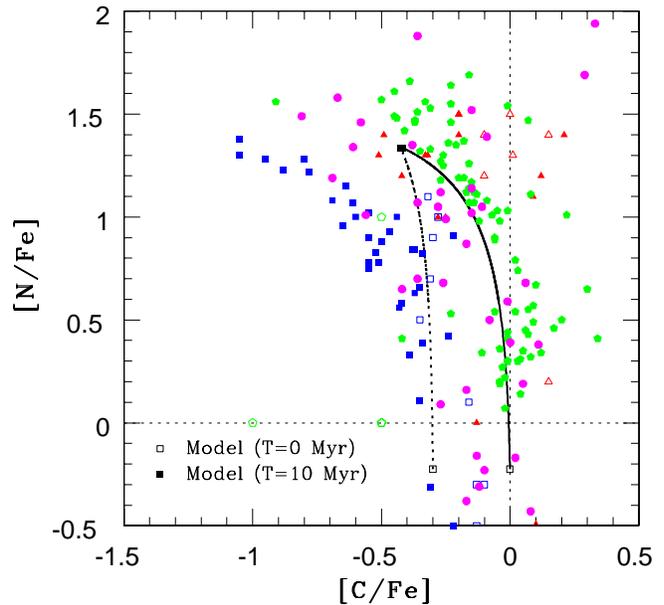,width=8.5cm}
\caption{ 
Comparison between models (M1 and M15) 
and observations  
for dwarfs and subgiants in NGC 6397  
(red open triangles; Carretta et al. 2005),
for those in NGC 6752  (red filled triangles; Carretta et al. 2005),
for those in 47 Tuc  (blue open squares; Carretta et al. 2005),
for SGB stars in M5  (blue filled squares; Cohen et al. 2002),
for main-sequence turn-off stars in  M13  
(green open pentagons; Briley et al. 2004a),
for MS stars in 47 Tuc  (green filled pentagons; Briley et al. 2004b),
and for subgiants and stars at the base of 
the red giant branch  in M15  (magenta filled circles; Cohen et al. 2005)
in  the [C/Fe]-[N/Fe] plane.
The time evolution of M1 and M15 are shown by
thick solid and dotted lines, respectively.
The typical error bars are $0.2$ for [C/Fe] and [N/Fe] 
(see Fig. 7 in Carretta et al. 2005).
}
\label{Figure. 14} 
\end{figure}

\section{Discussion}

\subsection{Comparison with observations}

It is more instructive for the present study to compare our model
predictions on abundance inhomogeneity within GCs with observations 
for GC MS
stars, rather than with the giant branch populations.  Comparison with
near-MS stars is to be preferred, because we do not have to consider
abundance changes due to deep mixing in evolved stars.
That said,  there are relatively few  studies  that show observational
data sets on abundances of MS  populations in GCs
(due to the difficulty in observing these
faint objects by the current generation of telescopes).
We thus use
available near-MS data sets in order to understand how well the ``pure'' AGB
pollution scenario can explain the observations for Galactic globular
clusters, and note that the data for evolved stars (i.e., RGB and AGB stars)
are not ideal for testing our models.
We here mainly focus on relations between [C/Fe] and [N/Fe]
using recent  observational data sets from Briley et al. (2004a, b),
Carretta et al. (2005), Cohen et al. (2002),
Cohen et al. (2005), and Smith et al. (2005).

Fig. 13 shows the comparison between the predicted
evolution of model in the [C/Fe]-[N/Fe] plane
and the observed [C/Fe] and [N/Fe] values for the red giants
in M13, M10, and NGC 7006 (Smith et al. 2005). 
It is clear from this figure
that (i) the observed anticorrelation between [C/Fe] and [N/Fe]
can be qualitatively reproduced,
(ii) the adopted ``pure'' AGB-pollution scenario, however,
can not explain the  observed smaller [C/Fe] ([C/Fe] $<-0.5$),
and (iii) it can not explain stars with high [N/Fe] ($>1.5$)
and low [C/Fe] ($<-0.5$).
These results clearly suggest that a combination of
deep mixing (e.g., see the new mechanism discussed by
Eggleton et al 2006) and primordial pollution from AGB stars in 
proto-GC clouds need to be considered for reasonable
explanations of the observations.

\begin{figure}
\psfig{file=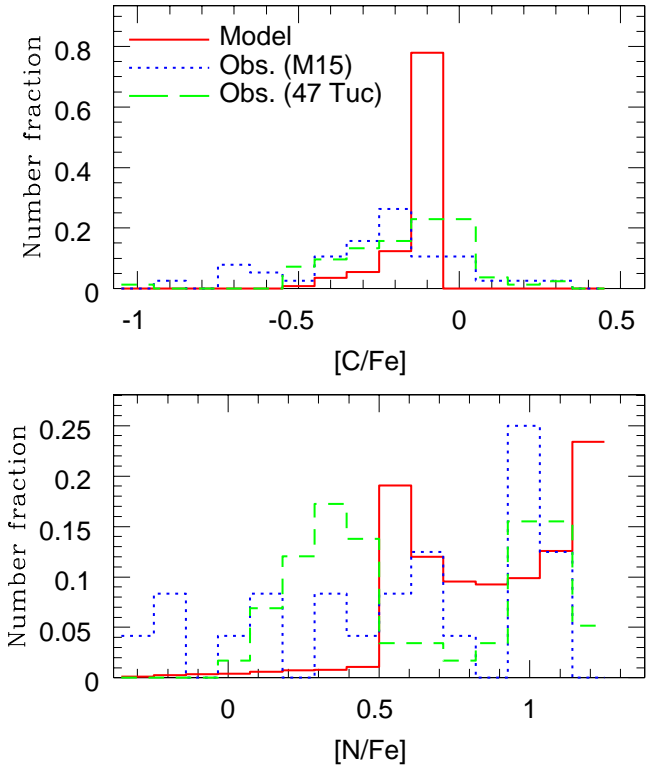,width=8.5cm}
\caption{ 
Comparison 
of [C/Fe] and [N/Fe] distributions 
between the model  M1 (red solid) 
and observations  
for subgiants and stars at the base of 
the red giant branch  in M15  (blue dotted; Cohen et al. 2005)
and for MS stars in 47 Tuc  (green dashed; Briley et al. 2004b).
}
\label{Figure. 15} 
\end{figure}

Fig. 14 shows the comparison between the predicted
evolution of the models on the [C/Fe]-[N/Fe] plane
and the observed [C/Fe] and [N/Fe] for less evolved
stars such as MS populations, subgiants, and dwarfs 
in NGC 6397  (Carretta et al. 2005), 
NGC 6752 (Carretta et al. 2005) , 47 Tuc (Briley et al. 2004b;
Carretta et al. 2005), M15 (Cohen et al. 2005), 
 M13 (Briley et al. 2004a), and M5 (Cohen et al. 2002).
Fig. 13 and Fig. 14 clearly indicate that
the models are much more consistent with observations
for less evolved stars (Fig. 14) than for evolved ones (Fig. 13),
though the models still have  difficulty in explaining
stars with low [C/Fe] ($<-0.5$).
Furthermore, the observed [C/Fe]-[N/Fe] anticorrelation 
and the apparently non-linear dependence of [N/Fe]  on [C/Fe]
are both consistent with the theoretical predictions.
These derived consistencies  may well provide support for
the adopted star formation histories of proto-GC clouds
in the external pollution scenario.

Fig. 15 shows the comparison between the predicted
[C/Fe] and [N/Fe] distributions  
and the observed ones for subgiants and stars
at the base of RGB in M15 and for MS stars in 47 Tuc.
It is clear from Fig. 15 that (i) both models and observations
show almost monomodal distributions in [C/Fe],
(ii) the predicted peak (in normalized number fraction)
in the [C/Fe] distribution is significantly higher than
those of the observations, (iii) the predicted bimodal
distribution is more similar to the observed one in 47 Tuc
than to the one in M15,
and (iv) the locations of the two peaks in [N/Fe] are 
however not exactly the same as the observed ones.
Thus the observed wider spread in [C/Fe] can not be
explained so well by the present models, though
the monomodal [C/Fe] and bimodal [N/Fe] distributions
are quite well reproduced by the models.

The present models also  predict the mean values of
[C/Fe]  and  [N/Fe]  in  GCs and thus can be compared with
the  ones derived from high-resolution spectroscopic
studies of   
{\it integrated spectra}
of GCs in nearby galaxies other than the Galaxy.
Ongoing and  future observational studies  based on large ground-based
telescopes with high-dispersion spectrographs 
enable us to derive [C/Fe]  and  [N/Fe] of GCs in other galaxies
by model fitting (e.g., Bernstein 2006).
Recent preliminary results reported by Strader et al. (2006)
for a  GC in M31 (034-096) show that [C/Fe] and [N/Fe] are roughly
0 dex and 0.6 dex, respectively, with the observational errors of
about 0.15 dex.
This result is more consistent with the present models with
smaller $s$ (that shows higher [C/Fe] and smaller [N/Fe]).
We suggest that a statistical study of
M31 GCs in terms of [C/Fe] and [N/Fe]  can provide new
insight into the differences in chemical pollution of
proto-GC clouds by AGB stars between M31 GCs and the Galactic ones.

\begin{figure}
\psfig{file=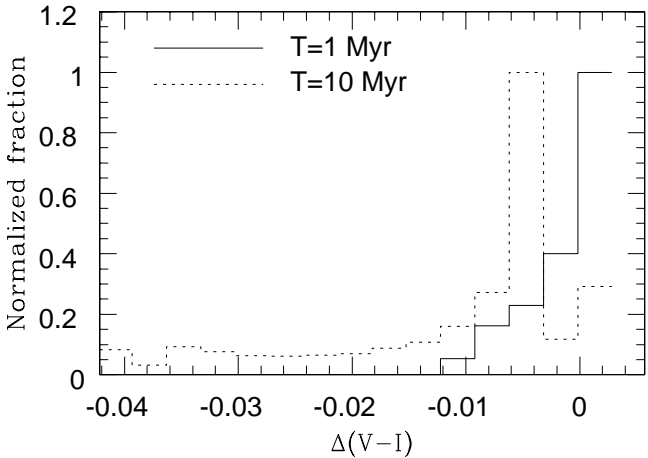,width=8.5cm}
\caption{ 
The predicted color spreads ($\Delta (V-I)$) 
due to Y inhomogeneity
for MS populations in M1 at T=1 Myr (solid) and 10 Myr (dotted).
$\Delta (V-I)$ is calculated from the results shown in Fig. 5
using the formula by D'Antona et al. 2005 (see the main text
for  details).
This figure implies that if GCs are formed from proto-GC 
clouds chemically polluted by AGB stars, they can show
color spreads (up to 0.05 mag in $V-I$) 
in their MS populations. 
}
\label{Figure. 15} 
\end{figure}

\subsection{Advantages and disadvantages of the external pollution
scenario}

\subsubsection{Advantages}

One of the serious problems in any self-pollution scenario is that
the number fraction of N-rich, C-depleted stars in a GC 
for a canonical IMF can not be
large enough to explain the observed fraction (e.g., Smith \& Norris 1982;
D'Antona \& Caloi 2004; Prantzsos
\&  Charbonnel
2006).  This problem can be solved if we assume
peculiar top-heavy IMFs (D'Antona \& Caloi 2004), for which GCs can
either be easily destroyed by the Galactic tidal field during
their evolution  (e.g., Chernoff \& Weinberg 1990) 
or be disintegrated quickly due to loss of a large
amount of mass through type-II supernovae explosion (Bekki \& Norris 2006).
Since both N-rich and N-normal components can be formed
from proto-GC clouds chemically polluted by field AGB stars 
(that do not become parts of GCs)
in the  external pollution scenario,
there are no serious problems on number fraction of N-rich, C-depleted
stars in the scenario with a canonical IMF.
Recently Pasquini et al. (2005) have suggested 
that the whole proto-GC cloud needs to be chemically enriched by a 
previous generation of stars based on the observational data
of the  Li abundance in 9 turn-off stars in NGC 6752: this is consistent
with the external pollution scenario.

GCs can be formed during or after the formation of the field stars
in dwarfs at very high $z$ in the external scenario.
A consequence of this is that 
the field stars in dwarfs on average should be older and more metal-poor
than GCs when GCs are formed
in this scenario.
Accordingly, if dwarfs with GCs can be destroyed by
the tidal field of the proto-Galaxy 
to form the Galactic field halo stars and the halo GCs 
soon after GC formation,
the field halo stars are highly likely to be more metal-poor
than the halo GCs in this scenario.
It is a well known observational  fact that 
the metallicities of the most metal-poor
halos stars are lower  than those of the most metal-poor
GCs in the Galaxy (e.g., Freeman 1993).
The earlier formation of field halo stars in comparison with GCs
in the external pollution scenario  
can naturally explain this observation.

Since proto-GC clouds are chemically  enriched by old field
stars (not by the very first stars like Pop III stars)
of low-mass dwarf galaxies
in the present models, 
the clouds  can  have  abundances similar to field stars 
(which can finally become halo stars) and be uniformly enriched
by $r-$ and $s-$process syntheses.
Therefore, GCs formed from these clouds 
can have abundances similar to those of field stars
except for light elements and accordingly
explain recent observations that have reported little  differences
in [Fe/H]-dependences
of [Sr/Fe], [Y/Fe], [Ba/Fe], and [Eu/Fe]
in main sequence turnoff stars and subgiants at the base of
the Red Giant Branch (RGB) in some  GCs (James et al. 2004).
Stars in the models with no third dredge-up keep
initial abundances of $s-$process elements
so that the models can be consistent with
the observed possible evidences for
homogeneous abundances of s-process elements in James et al. (2004).
The external pollution from AGB stars {\it with third dredge-up}
would be inconsistent with the above observations by 
James et al. (2004): future observations on
the degree of homogeneity in
the abundances of  $s-$process
elements  will give strong constrains on any AGB-pollution
scenario.

Recent abundance studies of the Galactic open cluster have
not yet revealed  possible abundance spreads  between stars
of these clusters (e.g., Friel et al. 2003; De Silva et al. 2006).
The external pollution scenario suggests that these clusters
are highly unlikely to show abundance inhomogeneities,
because they  are formed
from the Galactic gas clouds in which AGB pollution can not
proceed owing to very shallow gravitational potentials
of the clouds.
Hill et al. (2000) reported  that LMC clusters do not show
 Al-O anticorrelations,
which implies that AGB pollution did not proceed in these
systems. 
The external pollution scenario thus suggests that
these LMC clusters, which shows disky kinematics 
(e.g., Freeman et al. 1983), were formed not within dwarfs
that were the LMC's building blocks but within
the proto-LMC disk via a dissipative process at the 
formation epoch of the disk.
The observed differences in some elemental abundances 
between the LMC clusters and the Galactic GCs (Johnson et al. 2006)
may well be understood in terms of clusters originating
within {\it the forming disk of the LMC}.

\subsubsection{Disadvantages}

As pointed out by Fenner et al. (2004), chemical evolution
models of GCs based on AGB-pollution have some difficulties
in explaining quantitatively O-Na and Mg-Al anticorrelations
and correlations between $^{25}$Mg$/^{24}$Mg (and  $^{26}$Mg$/^{24}$Mg)
and light elements (e.g., [O/Fe] and [Na/Fe]).
Although some improvement over previous models can be seen
in the present models in which AGB stars experience
HBB but no third dredge-up,  the observed O-Na anticorrelation
can not be explained by them.
This incapability comes from the adopted stellar yield models
of AGB stars rather than from improper models for
early star formation histories within GCs. 
Given the fact that there are some uncertainties in model
calculations of yields in AGB stars (Ventura \& D'Antona 2005c;
Campbell et al 2006),
the above apparent failure does  not necessarily mean
that there is  a serious
problem with  the adopted AGB-pollution scenario.
Indeed, we remind the
reader that no consistent solution
currently exists. All proposals rely on
tweaking of some inputs, be it overshoot,
reaction rates or something else, or indeed
a combination. 
One solution for the above problem would be to adopt AGB models
with little third dredge-up (i.e., not the models with no dredge-up) 
in order to avoid unreasonably small values of [Na/Fe] in
the models. This contrived way of modeling, however,  should not
be regarded as a necessary  solution for this problem.
It may  say something about the characteristics of 
low-metallicity AGB stars or perhaps 
higher mass super-AGB stars -- maybe they do not 
suffer as much third dredge-up as expected by our models.

\subsection{The diversity in CN distributions}

Previous observational studies showed that there exists
variations in the strength of CN bands between the Galactic
GCs (e.g., Norris \& Smith 1981; Suntzeff 1981).
Norris (1988) investigated the value of the ``$r$'' parameter, 
which is defined as the ratio of the number of CN-strong to
CN-weak stars in a given system, for each of 12 GCs with 
$-1.9 \le {\rm [Fe/H]} \le -1.2$
and thereby found that the values of the $r$ parameter range from 0.22
(i.e., smaller fraction of CN-strong stars) to 3.29
(i.e., larger fraction).
Norris et al. (1984) also suggested that the observed  anticorrelation
between the behavior of the CN bands and that of the CH features in
GCs  (e.g.,  Norris \& Cottrell 1979;
Norris \& Freeman  1982;  Norris et al. 1984; Bell et al. 1983) 
is difficult to explain in terms of  the primordial 
pollution  (or self-pollution) scenarios.

We have demonstrated that (1) the distributions of N and C
are diverse depending on the two key parameters 
$s$ and ${\sigma}_{\rm I}$, 
(2) N distributions show
clear bimodality  in some models,
and (3) C and N abundances anticorrelate with each other
in the external pollution scenario.
These results  provide some physical basis for the origin 
of the observed diversity and for the observed C-N anticorrelation,
{\it if  we understand what determines the values of 
the above two key parameters
in proto-galaxies}.
Owing to the lack of extensive numerical simulations on
the dynamical fate of AGB ejecta in the central region
of proto-galaxies, 
it is unclear what physical processes in proto-galaxies
(e.g., the effectiveness of stellar feedback effects) can control
$s$ and ${\sigma}_{\rm I}$.
We plan to derive $s$ and ${\sigma}_{\rm I}$ and their dependences
on physical properties of proto-galaxies (e.g., masses and sizes)
in our future high-resolution simulations
and thereby discuss the origin of the observed diversity
of CN distributions in GC in the context of galaxy formation.

\subsection{Possible helium abundance inhomogeneity in GCs}

Recent photometric observations of stars in  $\omega$ Cen
have discovered a double main sequence
(DMS) in the color magnitude diagrams (CMDs) of its stellar populations
(Anderson 1997; Bedin et al. 2004).
One of the most promising interpretation is that
stars on the bluer main sequence (bMS) of the DMS
represents a very helium-rich ($Y\ge0.3$) population
(Norris 2004; Piotto et al. 2005).
Furthermore,
D'Antona et al. (2005) have reported 
helium abundance variation
among main sequence stars of NGC 2808.
Green (1980) first showed that there is a possible correlation
between the helium abundance of GCs and the locations of GCs with
respect to the Galactic center: inner GCs are more likely to
show higher helium abundance.
Recently Yong et al. (2005) have reported a possible trend
between [Fe/H] and [Al/Fe] for  giants stars
in NGC 6752 and thus suggested that
this trend can result from a possible correlation  
between the  H-He ratio and [Al/Fe],
which is a natural result of chemical  pollution by H-burning ejecta.
Although the total number of observations reporting
the possible abundance spread in helium is currently  very small,
these observations contain  valuable information 
not only on horizontal branch morphologies but also
on helium pollution processes in proto-GCs (e.g., D'Antona et al.
2005, 2006).

Important predictions of the present external pollution scenario 
on helium abundance properties of GCs  are 
(1) spread in helium abundance between stars,
(2) larger helium abundances in N-rich, C-depleted stars,
and (3) smaller helium abundance in GCs formed from gas
less polluted by field AGB stars.
Furthermore, we can provide some prediction on color  distributions
of stars  for the main-sequence (MS) populations of GCs
by using the model developed by D'Antona et al. (2005).
Differences in Y between
stars with normal Y (=0.24) and those with higher/lower Y
can be  converted into
color difference ($\Delta (V-I)$) by 
using the following formula (D'Antona et al. 2005):
\begin{equation}
\frac{\Delta(V-I)}{\Delta Y} \simeq -0.438.
\end{equation}
Fig. 16 clearly demonstrates that  the color distributions of stars in the
fiducial model M1 shows significant dispersions in $V-I$ of the MS. 
This dispersion can be observed as spreads in the CM relations of
MS populations of GCs and thus as evidence for GC formation
from AGB ejecta.
 
Several authors have pointed out that He abundance inhomogeneity
of stars in GCs can be imprinted on photometric properties of stellar
populations on the horizontal branch (HB) of  GCs
(e.g., Norris 2004; D'Antona et al. 2006; Lee et al. 2005).
These authors suggested that a GC with a He-rich population
can (e.g., NGC 2808) show 
a characteristic HB color and luminosity distribution,
a bimodal HB distribution, and faint, very hot 
HB stars.
Since a GC with a larger fraction of N-rich stars
can have a  higher degree of He abundance inhomogeneity in
the external pollution scenario,
the HB properties of GCs can be correlated with
the fraction of N-rich stars (i.e., $r$ parameter).
If the origin of the second parameter problem is related to
the difference in helium abundance between GCs
(see Catlan 2005 for a recent review),
the  external pollution scenario suggests that the origin
of the problem can be  closely associated with 
differences in AGB pollution processes  (controlled mainly by
the two parameters, $s$ and ${\sigma}_{\rm I}$)
between different proto-GC clouds in the central regions
of forming dwarfs at high $z$.

%\subsection{No or weak 3rd dredge-up in very metal-poor AGB stars}

\subsection{Origin of the observed correlation between
the number fraction of CN-strong populations  and  ellipticities of GCs}

Norris (1987) investigated possible correlations between
the $r$ parameter 
and other physical properties of GCs
(e.g., metallicities and luminosities) and found that
there exists  a  correlation between
ellipticities ($\epsilon$) of GC shapes and the values of the $r$ parameters
(See Smith (2002) for recent confirmation of this correlation).
This result means that if the flattened shapes are due to rotational
kinematics of GCs (rather than anisotropy of velocity dispersion),
GCs with a larger amount of angular momentum
are more likely to show larger $r$.
Norris (1987) suggested that {\it if global rotation of 
a GC positively correlate with rotation of individual stars within
the GC and if evolutionary mixing  responsible for the CN abundance
inhomogeneity is driven by stellar rotation},  the observed correlation
can be readily explained.
Norris (1987) however pointed out that this explanation has a difficulty 
in explaining  the existence of CN-enhanced stars near the main-sequence
turnoff of 47 Tuc (Bell et al. 1983) 
unless the effects of rotation are efficient
for main-sequence stars.
Furthermore it is  unclear 
why stars within GCs with a larger amount of intrinsic
angular momentum would 
have a larger amount of stellar rotation in this scenario
by Norris (1987).

In the present external pollution scenario,  
the values of the $r$ parameters depend
strongly on the relative mass fraction of AGB ejecta
with respect to the total gas mass (i.e., AGB ejecta + fresh infall gas)
in proto-GCs: The higher the fraction is, the larger the $r$ becomes.
One of possible explanations for the above $\epsilon$-$r$ correlation
in the context of the external pollution
is therefore that whenever  a larger amount of AGB
ejecta can be transferred into proto-GC regions (i.e., central regions
of their host dwarfs), the gas is highly
likely to  have a larger amount of
angular momentum, from which GCs with a larger amount of angular momentum
and thus with larger $\epsilon$ can be formed.
We suggest that the above situation of the larger amount of
AGB ejecta  with higher (specific) angular momentum 
is likely if field stars in the central regions of dwarfs
(thus, AGB polluters for proto-GCs) have rotational kinematics:
AGB ejecta from such stars rotating dwarfs can also
have orbital angular momentum (with respect to the centers of dwarfs) 
so that GCs formed from the ejecta in the centers
can have rotational kinematics and thus large  $\epsilon$
after conversion of orbital angular momentum of gas into intrinsic
spin angular momentum of GCs during GC formation.

It is thus our future study to  investigate (i) whether 
orbital angular momentum of AGB ejecta 
from field stars rotating dwarfs can be really converted
into intrinsic spin angular momentum of GCs and (ii) whether
GCs have a larger 
amount of intrinsic spin angular momentum thus larger  $\epsilon$,
if they are formed from gas with a larger fraction of AGB ejecta.
High-resolution numerical simulations on stellar and gas dynamics in
the central $0.1-100$pc of dwarfs at high $z$ would be  essential
for this investigation.
We plan to perform such chemodynamical simulations with AGB feedback
effects (e.g., thermal heating and  return of AGB ejecta to ISM)
to understand the origin of the observed $\epsilon$-$r$ correlation
in the Galactic GCs. 
Nuclear dynamics in proto-galaxies would be dependent 
on the depth of their gravitational potential wells  thus
on their masses and sizes.
Our future simulations will therefore enable us to understand
whether very flattened GCs (e.g., $\omega$ Cen) can be formed
in less or more massive dwarfs  at high $z$.
%Carretta (2005) has discovered a new correlation between
%the degree of abundance inhomogeneity and the orbital parameters
%of GCs.

\subsection{Origin of possible higher nitrogen abundances 
for some bright  GCs in M31}

Recently, Li \& Burstein (2003) have found that the NH absorption line
is far stronger in (three) metal-rich GCs of M31 than it is
in the Galactic GCs at a given value of CH or [Fe/H].
Beasley et al. (2004) also have revealed that the near-UV 
cyanogen features of M31 GCs are  strongly enhanced with respect
to the Galactic GCs for $-1.5 < {\rm [Fe/H]} < -0.3$.
These observations imply that [N/Fe] can be higher in
some M31 GCs than in the Galactic ones for a same metallicity range.
It has been also revealed that M31 GCs having very strong NH
absorptions  are more  luminous with
$-11 < M_{\rm V} < -8.5$ (e.g., Burstein et al. 2004).
This implies that the possibly higher [N/Fe] is due to 
a larger number of bright GCs in M31, which can have the higher values 
for some physical reasons.
It remains,  however, totally unclear why these bright M31 GCs have
high [N/Fe].

The present models have shown  that mean values of [N/Fe] of GCs,
which should  be compared with the [N/Fe] estimated from the
integrated spectra of M31 GCs, can be as high as 0.9. 
This result, combined with recent theoretical results and
observational ones on nucleus formation of galaxies
(e.g., Bekki 2006; Bekki et al. 2006;  C\^ote et al. 2006),
can provide the following  answer for the above question.
Based on analytical calculations, Bekki (2006)  pointed out
that AGB ejecta can be more effectively trapped in nuclear
region of more massive dwarfs embedded in dark matter halos
so that AGB pollution can proceeds to a greater extent in these
dwarfs. Inward radial gas-transfer into nuclear regions, which is indispensable
for the formation of compact stellar systems, is demonstrated to
be more efficient in more massive dwarfs (Bekki et al. 2006).
Photometric studies on nuclear properties of dwarfs
based on the ACS Virgo Cluster Survey
have recently revealed that stellar nuclei are more massive
in more massive dwarfs (C\^ote et al. 2006). 
Therefore we can claim that if some bright
GCs in M31 originate from nuclei of brighter nucleated dwarfs,
the observed higher [N/Fe] can be naturally explained.

Thus the origin of the observed possible differences in [N/Fe] between
M31 GCs and the Galactic ones could reflect the fact that
the proto-M31 contained a larger number of brighter dwarfs where
AGB pollution could proceed more effectively in their central regions to form
GCs with higher [N/Fe].
These more massive dwarfs were destroyed to form the M31's
stellar halo during the hierarchical
formation of M31 through merging of these dwarfs.
Because of the mass-metallicity relation of field stellar
populations of dwarfs (e.g. Durrell et al. 1996), 
the developed stellar halo can show a higher metallicity
and is consistent with observations (e.g., Reitzel et al. 1998).
If [N/Fe] of GCs can be controlled by nuclear dynamics
of their host dwarfs at very high $z$,
as suggested by the present study,
more luminous galaxies would  have both
a larger number of GCs with high  [N/Fe]
and more metal-rich (thus redder) stellar halos. 
We thus suggest that future high-resolution spectroscopic
studies for GC abundances in galaxies
beyond the Local Group, combined with wide-field imaging
of stellar halos in these galaxies, 
can give some constraints on the formation histories of
GCs within dwarfs at very  high $z$.

\section{Conclusions}

We have investigated the abundance inhomogeneity  among  the
light elements (e.g., C, N, O, Na, and Al) in stars in
the Galactic  GCs 
by adopting the external pollution scenario 
and using latest stellar yield models of metal-poor AGB stars
with and without third dredge-up.
In the  external pollution scenario,
GCs within a forming low-mass dwarf embedded in a dark matter halo
can be formed from the mixture of (i) gas ejected from 
the field
AGB stars formed earlier in the dwarf 
and (ii) the interstellar gas infalling to the central
region of the dwarf. 
%Since the stellar yield from models without third dredge-up can much
%better explain the observed properties of GCs,
We extensively investigated our models
with no third dredge-up  over  wide parameter ranges.
We summarise our principal results 
of the models as follows.

(1) The ratio of the total mass of infalling gas
to that of AGB ejecta during GC formation within a forming dwarf galaxy
(described by  the  $s$ parameter) and the time scale of 
gas infall (${\sigma}_{\rm I}$) are the most important  parameters  
that can determine abundance properties of GCs.

(2) Both [N/Fe] and [C/Fe] can be diverse among stars
within a GC owing to chemical pollution from field AGB stars. 
[N/Fe] distributions in some GCs can clearly show 
bimodality wheres [C/Fe] can be  monomodal in most models.
[N/Fe] distributions depend on $s$,  such that
models with smaller $s$ (i.e., larger mass fraction
of AGB ejecta used for GC formation)  show 
the [N/Fe] bimodality more clearly.

(3) N-rich, C-poor stars in GCs also have higher He abundances
owing to pollution from massive AGB stars with He-rich ejecta.
The number fraction of He-rich stars (Y $>0.30$) is higher
for the models with smaller $s$ 
and shorter ${\sigma}_{\rm I}$
for $3\le s \le24$ and $10^5 \le {\sigma}_{\rm I} \le 10^7$ yr.
He abundances of stars can correlate with [N/Fe] and [Al/Fe]
and anticorrelate with [C/Fe],  [O/Fe], and  [Na/Fe] 
within GCs (and a simple consequence of  H-burning).

(4) Although our models can much better explain the observed
C-N and Mg-Al anticorrelations than previous theoretical models,
the observed O-Na anticorrelation
can not by simply explained by any models with different parameters
in this scenario.
The inability of the present models 
to match the observations results from
the adopted AGB yields and thus suggests
that models using different AGB yields need to be explored  
to assess the viability of  AGB pollution scenarios
(For more details, see  Appendix B).

(5) Extensive comparison between the models and the observations
demonstrates that the predicted [C/Fe]-[N/Fe] anticorrelation can be 
more quantitatively consistent with observations for 
less evolved stars (MS populations, subgiants, and dwarfs)
than for evolved ones (e.g., red giants).
The comparison also demonstrates that the observed [C/Fe]
and [N/Fe] distributions for less evolved stars
are at least qualitatively consistent with the model predictions.

(6) One important prediction of our models is
that GCs with higher degrees of abundance inhomogeneity
should show a larger color spread (e.g., $\Delta (V-I)$)
in their MS populations as the result of  
bluer stars formed from He-rich gas.
Accordingly, it would be  worthwhile for future
observational studies to investigate correlations
between the $r$ parameters (i.e., the number ratio of CN-strong to CN-weak
stars) and $\Delta (V-I)$ in GCs.

(7) The values of the two key parameters ($s$ and ${\sigma}_{\rm I}$)
are suggested to be controlled by stellar and gas dynamics
in the central regions of  
host galaxies of GCs. 
Central stellar and gas dynamics 
(e.g., gas fueling rates to the central 10 pc) have been suggested  to be
controlled by global properties (e.g., total masses and densities)
of galaxies.
Accordingly,  abundance inhomogeneities of GCs
may  have fossil information on  (i) formation and evolution histories
of central regions of defunct dwarfs and (ii) physical properties
of the dwarfs.

Thus we have shown that 
the CN-bimodality and the number fraction of C-depleted, N-rich
populations of a  GC
can give strong constants on the short-term (an order of $10^7$ yr)
star formation history of the proto-GC cloud of the GC,
if the cloud is chemically polluted by AGB stars.
We accordingly suggest that 
any theoretical models 
for the origin of GC abundances   
need to explain 
not only the O-Na and Mg-Al anticorrelations
but also the bimodal CN abundance distributions in GCs.

\section*{Acknowledgments}
We are  grateful to the anonymous referee for valuable comments,
which contribute to improve the present paper.
K.B., J.C.L., and J.E.N. 
acknowledge the financial support of the Australian Research 
Council throughout the course of this work.
SWC would like to thank the Australian Partnership for Advanced Computing
(APAC) for the computing time granted under project g61, "Element Production by Intermediate Mass Red-Giant Stars".

\appendix
\section{A possible model for the bMS in $\omega$  Cen}

\begin{figure}
\psfig{file=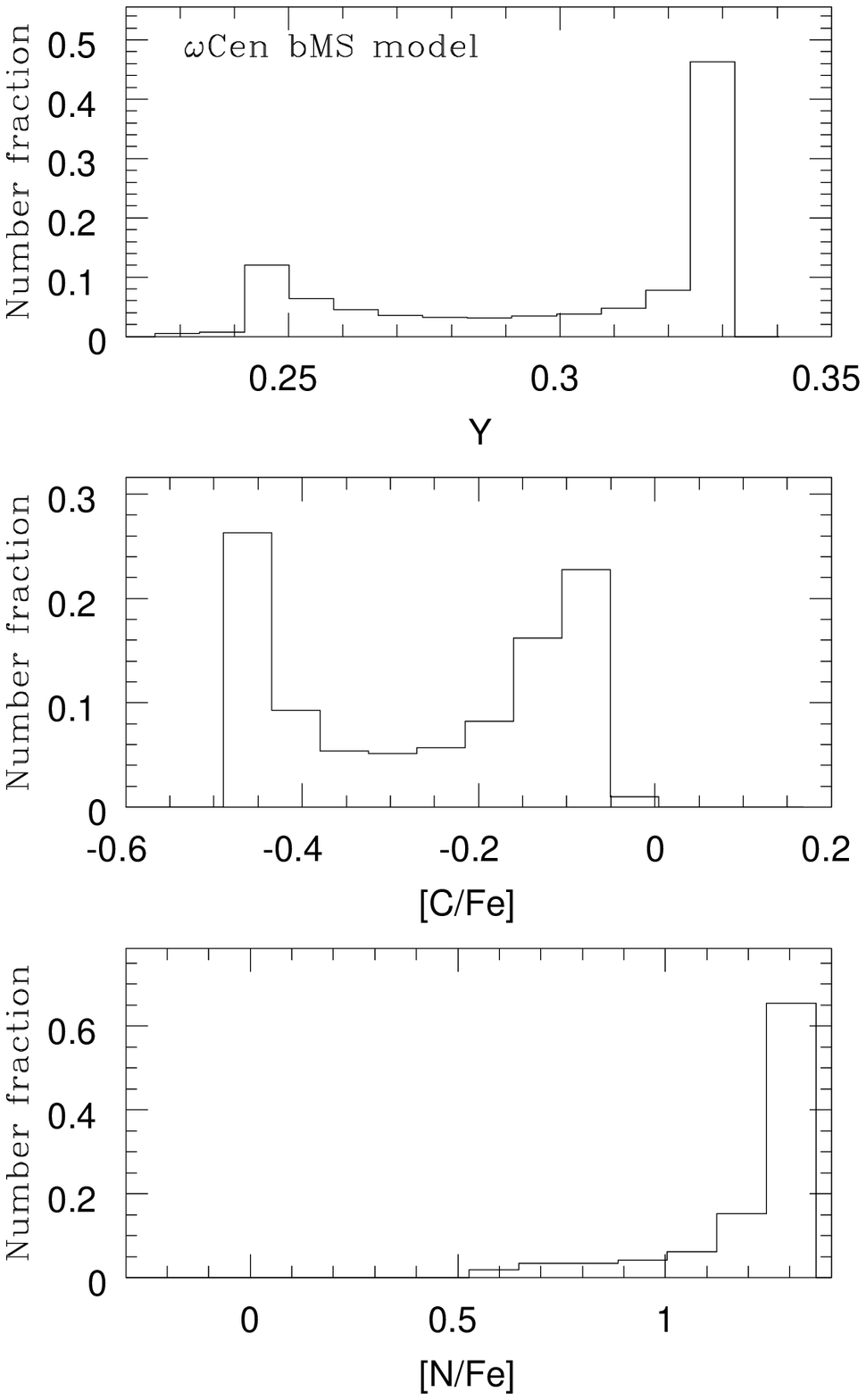,width=8.5cm}
\caption{ 
A possible model of Y (top), [C/Fe] (middle), and [N/Fe] (bottom)
distributions for stars on the blue main sequence
(bMS) in  $\omega$ Cen  which is suggested to have
a significant fraction of He-rich stars in the bMS population.  
}
\label{Figure. A1} 
\end{figure}

Bedin et al. (2004) suggested  that   
stars on the bluer main sequence (bMS) of the  double
main sequence (DMS)
can be a very helium-rich ($Y\ge0.3$) population.
A number of recent investigations have suggested that 
a higher degree of He enhancement in the bMS
is a viable explanation for the origin of the DMS
(e.g., Lee et al. 2005; Norris 2004; Piotto et al. 2005)
and recent analytical works by Bekki \& Norris (2006)
have suggested that He enhancement due to self-pollution 
(or more correctly, external pollution) of AGB stars can 
explain the origin of the bMS. 
It however remains unclear whether such AGB pollution
scenarios can {\it quantitatively} explain the observed
abundance properties (e.g., [C/Fe], [N/Fe], and Y) 
in a self-consistent manner.

Fig. A1 shows Y, [C/Fe], and [N/Fe] distributions of
the model M14 in which very small $s$ (i.e., a higher
degree of AGB pollution), smaller ${\sigma}_{\rm I}$
(i.e., more rapid gas infall), and smaller $M_{\rm g}(0)$
(i.e., more rapid chemical enrichment) are adopted
so that the model shows a large fraction of He-rich stars.
Owing to the modeled very larger degree of AGB pollution,
stars in this model shows a large fraction of 
He-rich ($>0.3$) and N-rich ($>1$) stars.
Piotto et al. (2005) have reported  that (i) the He-rich bMS
shows a rather large nitrogen abundance ([N/M] $\sim 1-1.5$)
and (ii) it is not very rich in C ([C/M]=0).
The result in Fig. A1 implies that these observations  can
be due to AGB pollution, though the C abundance of the bMS
is not so consistent with the predicted one.
An important result in Fig. A1 is that 
some fraction ($\sim 0.1$) of stars show relatively normal
Y ($0.24-0.25$). This suggest that if the bMS is formed
from gas polluted by AGB stars,
stars with relatively normal Y are  also formed,
though the mass fraction of the He-normal population
is small.
This may well provide a clue to the origin of the
observed normal Y (=0.246) for the metal-intermediate
RRL stars (that would have been formed
with the bMS) in $\omega$ Cen (Sollima et al. 2006).

Very luminous GCs such as $\omega$ Cen in the Galaxy and G1 
in M31 have long been
suggested to originate from nuclei of ancient nucleated systems,
where continuous gas infall and chemical enrichment were
highly likely to occur (e.g., Hilker \& Richtler 2000; 
Bekki \& Freeman  2003; Bekki \& Chiba 2004).
As demonstrated above,  the total mass of AGB ejecta used
for the He-rich population in $\omega$ Cen needs to be $\sim$ 6
times larger than that of infalling gas not enriched by AGB
stars. It remains unclear whether this can really happen
in the nuclei of dwarf galaxies that are considered to
be progenitors of very luminous GCs like $\omega$ Cen and G1.

%\begin{figure}
%\psfig{file=f16.eps,width=8.5cm}
%\caption{ 
%A possible model of Y (top), [C/Fe] (middle), and [N/Fe] (bottom)
%distributions for NGC 6752 for which the number fraction of
%CN-strong stars are similar to that of CN-weak ones. 
%}
%\label{Figure. A1} 
%\end{figure}

\section{Comments on AGB models with and without third dredge-up}

Current standard models of intermediate-mass AGB stars with third dredge-up 
(3DUP) do not reproduce the observed C-N, O-Na or Mg-Al inverse-correlations 
observed in most GCs. We note however that the models of Ventura \& D'Antona (2005b) do 
come close, due to their use and development of an alternative theory of 
convection, which alters the evolution of their models significantly 
(see e.g., Ventura \& D'Antona 2005a). In the standard models the combination of hot 
bottom burning (HBB) and 3DUP during AGB evolution does predict enhancements 
of nitrogen, sodium and aluminium, and a reduction in oxygen -- all of which 
are needed to explain the observations -- however they also predict an increase 
of carbon and magnesium, giving positive C-N and Mg-Al correlations which is 
in direct violation of the observations. In addition to this the models do not 
\emph{quantitatively} account for the large degree of O depletion and tend to 
over-produce Na, thus falling short of the mark with the O-Na 
inverse-correlation as well (e.g., Denissenkov et al. 1997; Denissenkov \& 
Herwig 2003; Ventura et al. 2004). Furthermore the standard AGB models produce 
ejecta with an overall increase in CNO nuclei of $\sim$ 1 to 3 dex (mainly due 
to the dredge-up of carbon), whilst observations show that the sum of CNO 
nuclei is virtually constant, varying by a factor of only two or so at most 
(e.g., Pilachowski 1988; Smith et al. 1996; Ivans et al. 1999). These 
discrepancies are based on confronting single stellar models with observations. 
Taking the evolution of an entire cluster of stars into account -- in 
particular convolving with an IMF -- presents further problems as the stars 
with the prerequisite hot hydrogen burning are few in number when using a 
standard IMF (e.g., Fenner et al. 2004), whilst these stars need to contribute 
an amount of polluting material that is a large fraction of the GC mass. 
Despite all these discrepancies it should be stressed that the AGB pollution 
scenario should not be dismissed outright just yet, as the models are well 
known to have many serious uncertainties (e.g., the treatment of convection, 
overshoot, mass loss and the uncertainties in reaction rates). For example 
the non-standard models of Ventura \& D'Antona (2005b), which use a different 
convection model, can reproduce most of the abundance anomalies, although 
some overshooting and the adoption of a high mass-loss rate is required. In 
addition to the uncertainties in the input physics of all stellar models, 
numerical details can also significantly affect the results. For example, 
time-stepping during diffusive 
mixing can cause changes in the stellar structure, hence altering the temperature 
at the bottom of the convective envelope, and thus the resulting nucleosynthesis 
and yields (Siess, private communication). Attempts have been made to reconcile 
the (standard) models with the observations but have universally ended with a 
need to `tweak' the relative effects of HBB and 3DUP to an unpalatable degree 
(e.g., Denissenkov \& Herwig 2003), also suggested by Fenner et al. (2004). 
The models using a different convection formalism (Ventura \& D'Antona 2005b) 
show a better agreement with observations but, as mentioned earlier, also need 
some overshooting and the adoption of a high mass-loss rate.

With the results of all these stellar modelling studies in mind we have chosen 
to investigate the limiting case by utilising yields from AGB models that do not 
experience any 3DUP, which we have calculated ourselves. Although `turning off' 
3DUP in our models is ad-hoc, it was surprisingly 
easy to achieve. By adding an extra physical condition for the determination of 
convective boundaries (the molecular weight gradient) via implementing the Ledoux 
criterion rather than the Schwarzschild criterion we found that no dredge-up at 
all occurred in our models. We note that we have not extended the convective 
boundaries past the formal ones given by the Ledoux criterion. We leave the 
detailed description of the models for a separate paper, but now discuss the 
nucleosynthetic results relevant for the current problem.

Decoupling 3DUP from HBB exposes the dependence of the yields on the interplay 
between the two processes and provides a possible solution to the mismatch 
between theory and observation. At the very least, by separating the effects it 
is a useful investigative tool. A salient result from our new stellar models, 
evident in Table 1, is that oxygen is now heavily depleted in the more massive 
stars. Also of importance is  the fact that there is now an opposite trend 
for Mg (in the same stars) -- it is depleted instead of produced. Furthermore, 
the C is depleted whilst the N yield remains high. All of these features derive 
from the lack of 3DUP, as O is now not periodically replenished, C can not 
increase and fresh fuel for the MgAl cycle is not available. The lack of 3DUP 
also implies that the sum of CNO nuclei will remain constant, as fresh CNO is 
not added to the envelope and the CNO burning cycles (occurring at the base of 
the convective envelope) conserve the number of CNO nuclei. All these features 
indicate that a better fit to the observations is expected (IMF withstanding). 
Also, as stated above, these models may simulate the effects of a previously 
ignored population of stars, the Super-AGB stars. The main discrepancy we now 
have -- in terms of comparing at a single mass -- is with Na. At the high 
masses (ie. temperatures) required to deplete O and Mg via the MgAl and ON 
cycles, Na is not produced enough, or rather it is destroyed too much by 
proton captures, resulting in low yields. Moving to lower masses (temperatures) 
improves the Na situation but prevents O and Mg from being depleted. 
Interestingly this is the identical problem that Ventura \& D'Antona (2005b) 
have with their best-fit (for the Na-O inverse-correlation) model. 
Denissenkov \& Herwig (2003) also report this problem. We note that the models 
of Ventura \& D'Antona (2005b) are quite independent as they use a different 
treatment for convection. In a recent and very timely study 
Ventura \& D'Antona (2006) have investigated this problem in detail, finding 
that their stellar model requires either the addition of some overshooting or 
a reduction of the $^{23}$Na$(p,\alpha)^{20}$Ne reaction rate in order to 
produce the required Na yield whilst maintaining the O destruction. The Ventura 
and D'Antona (2005b, 2006) models, like our more ad-hoc models, otherwise match 
the abundance anomalies fairly well, with the strong exception of the Mg isotopes, 
which seem resistant to any approach tried thus far. Note that we do not suggest 
that real AGB models of low metallicity do not suffer from the 3DUP. Rather we have
 used this as an investigative tool to simulate reduced 3DUP, or SAGB stars 
with little or no 3DUP, or some other source of (almost) pure H-burning. Clearly 
a thorough understanding, and reliable quantitative prediction of 3DUP is still
 required, for stars of all masses and metallicities.

\end{document}